\documentclass[%
 reprint,
 amsmath,amssymb,
 aps,
floatfix,
]{revtex4-2}

\usepackage[T1]{fontenc}

\usepackage{hyperref}
\usepackage{graphicx}	%
\usepackage{amsmath}	%
\usepackage{subcaption}
\usepackage{xcolor}
\usepackage{booktabs}
\usepackage{adjustbox}
\usepackage{makecell} %

\newcommand{\horres}[2]{#1\,#2\,$\times$\,#1\,#2}

\newcommand{\revM}[1]{#1}

\begin{document}
\title{OTProf: estimating high-resolution profiles of optical turbulence ($C_n^2$) from reanalysis using deep learning}

\author{Maximilian Pierzyna}
\email{m.pierzyna@tudelft.nl}
\affiliation{Department of Geoscience and Remote Sensing, Delft University of Technology, Delft, The Netherlands}

\author{Sukanta Basu}
\affiliation{Atmospheric Sciences Research Center, University at Albany, Albany, USA}
\affiliation{Department of Environmental and Sustainable Engineering, University at Albany, Albany, USA}

\author{Rudolf Saathof}
\affiliation{Faculty of Aerospace Engineering, Delft University of Technology, Delft, The Netherlands}

\date{\today}

\begin{abstract}
    Accurate, high-resolution vertical profiles of optical turbulence ($C_n^2$), which reflect local meteorology and topography, are crucial for ground-based optical astronomy and free-space optical communication. 
    However, measuring these profiles or generating them with numerical weather models requires substantial operational or computational effort. 
    In this work, we present OTProf, a deep-learning method that estimates high-resolution $C_n^2$ profiles from widely available coarse-resolution ERA5 reanalysis data. 
    OTProf is based on the Squeezeformer architecture, modified for super-resolution by introducing grid-agnostic interpolation in the latent space. 
    We train and evaluate OTProf in the Netherlands using high-resolution $C_n^2$ data obtained from the Weather Research and Forecasting (WRF) model. 
    The commonly used Hufnagel-Valley model serves as the baseline. 
    Overall, OTProf reproduces the vertical structure of WRF-based $C_n^2$ more accurately than Hufnagel-Valley and yields more accurate estimates of the Fried parameter $r_0$ and the scintillation index $\sigma_I^2$. 
    Out-of-domain performance is assessed using a smaller WRF-based $C_n^2$ dataset covering the Canary Islands. 
    Initial results indicate a performance drop because the orography differs markedly from the training domain.
    We show that output bias correction based on a few local simulations can help OTProf to adapt, but more work for domain adaptation is needed. 
    OTProf is therefore primarily recommended for in-domain applications, e.g., generating $C_n^2$ climatologies from long ERA5 records.
\end{abstract}

\maketitle

\section{Introduction}
Vertical profiles of optical turbulence (OT), quantified by the refractive index structure parameter $C_n^2$, are essential for characterizing the performance of ground-based astronomical observations or free-space optical communication (FSOC) systems.
These profiles determine key parameters such as the Fried parameter $r_0$ and the scintillation index $\sigma_I^2$, which directly impact beam spreading, wavefront distortion, and signal fading (\citet{andrews2005}, \citet{kaushal2017}).
To predict optical turbulence conditions, the community has traditionally relied on empirical models such as the non-parametric SLC models \citep{miller1979}, the parametric Hufnagel-Valley (HV) model \citep{hufnagel1974,valley1980,ulrich1988}, and others (cf.~\citet{good1988} or \citet{beland1993} for an overview).
However, these models have a limited ability to account for local meteorology.
The SLC models, for example, are non-parametric, so their $C_n^2$ estimates do not depend on location, time, or local meteorology at all.
The HV model has a weak dependence on local meteorology but still prescribes a largely generic vertical structure.
Efforts are still being made to better adjust HV-based models to local boundary layer conditions (e.g., \citet{comeron2005,andrews2009,stotts2023,dasgupta2026}), but the generic exponential shape of the models remains a fundamental limitation.
Alternatively, numerical weather prediction models like the Weather Research and Forecasting (WRF) model \citep{skamarock2021} can provide physically consistent high-resolution $C_n^2$ profiles, which do account for local conditions (e.g., \citet{masciadri1999}, \citet{cherubini2013}, \citet{giordano2014,Basu2020,rafalimanana2022}), but they are computationally expensive to obtain.
This cost makes it impractical to simulate the multi-year periods needed for regional $C_n^2$ climatologies, e.g., for site selection or link availability studies.
Given these limitations, more efficient, physically consistent approaches are needed to predict optical turbulence profiles from meteorological data.

This study addresses this need by proposing OTProf, a machine learning approach to estimate high-resolution (HR) vertical $C_n^2$ profiles from coarse, low-resolution (LR) meteorological data, such as the globally available ERA5 reanalysis (the fifth-generation atmospheric reanalysis of the European Centre for Medium-Range Weather Forecasts, ECMWF).
Our approach requires solving two tasks simultaneously: regression and super-resolution.
The regression task is to predict optical turbulence ($C_n^2$) from meteorological variables. 
At the same time, super-resolution aims to achieve a vertical output resolution of the predicted profiles that exceeds the vertical input resolution of the meteorological data.
In OTProf, this super-resolution is realized by interpolating in the latent feature space of the model rather than on the input or output profiles directly.
The Netherlands (NL) are selected as an example study area, for which we generate a year-long high-resolution WRF model-based training dataset.
Once trained, OTProf can be applied to long ERA5 records, e.g., to extend a limited set of WRF simulations to a regional $C_n^2$ climatology.
The ERA5 reanalysis \citep{era5} on pressure levels (PL), denoted ERA5-PL, is used as LR meteorological input data.
ERA5 and WRF differ in resolution and physics parameterizations, resulting in misaligned datasets, which complicate deep learning (DL) training. 
To address this challenge, OTProf uses a decoupled training approach followed by output bias correction.
We benchmark OTProf on experiments of increasing complexity and compare it against the HV model as the baseline.
To test the spatial generalization, we additionally apply the OTProf model trained on the Netherlands to the Canary Islands, which differ strongly in climate and orography.

Throughout this study, the reference $C_n^2$ profiles are derived from WRF, so OTProf is trained and evaluated to reproduce model-based optical turbulence rather than observations.
Observations are typically less smooth than WRF-based $C_n^2$ estimates, so OTProf (like WRF) is expected to yield larger errors when confronted with observations.
Because our proposed latent-space interpolation for super-resolution is grid-agnostic, we recommend future research to fine-tune OTProf on observed profiles to incorporate more fine-scale features.
For the current study, we consider mesoscale-model-based $C_n^2$ data as well suited because of their wide acceptance in the community (e.g., \citet{masciadri1999,masciadri2013,masciadri2020,Basu2020,pierzyna2024}).

The manuscript is organized as follows.
In sec.~\ref{sec:methods}, we formalize the learning problem and introduce the DL architecture and the performance evaluation strategy.
The LR and HR datasets, as well as the DL dataset generation, are introduced in sec.~\ref{sec:datasets}.
Section~\ref{sec:results} compares OTProf against the HV model, considering both the performance to estimate $C_n^2$ profiles and the two integrated parameters $r_0$ and $\sigma_I^2$, and assesses the transfer to the Canary Islands.
Finally, the study is concluded in sec.~\ref{sec:conclusion}.

\section{Methods}\label{sec:methods}
\begin{figure}[t]
   \centering
   \begin{subfigure}[t]{.8\columnwidth}
      \caption{ERA5 \horres{0.25}{$^\circ$}}
      \includegraphics[width=\textwidth,trim=0mm 0mm 99mm 0mm,clip]{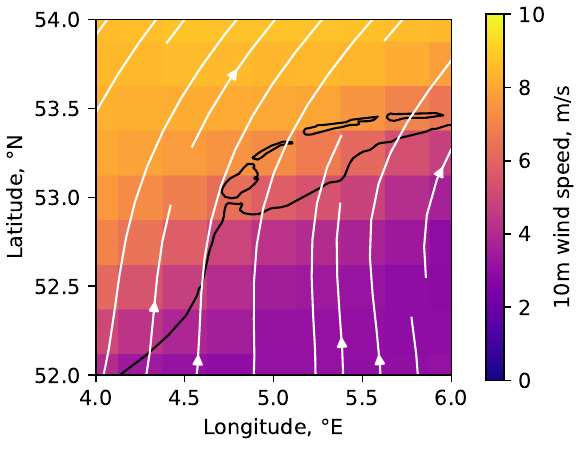}
   \end{subfigure}
   \begin{subfigure}[t]{.8\columnwidth}
      \caption{WRF \horres{2}{km}}
      \includegraphics[width=\textwidth,trim=99mm 0mm 0mm 0mm,clip]{figures/wrf_era5_wind.pdf}
   \end{subfigure}
   \caption{Example of mismatching 10\,m wind fields due to higher spatial resolution of WRF compared to ERA5.}%
   \label{fig:res_diff}
\end{figure}

\begin{figure*}[t]
   \centering
   \includegraphics[width=\textwidth,trim=7mm 6mm 30mm 6mm,clip]{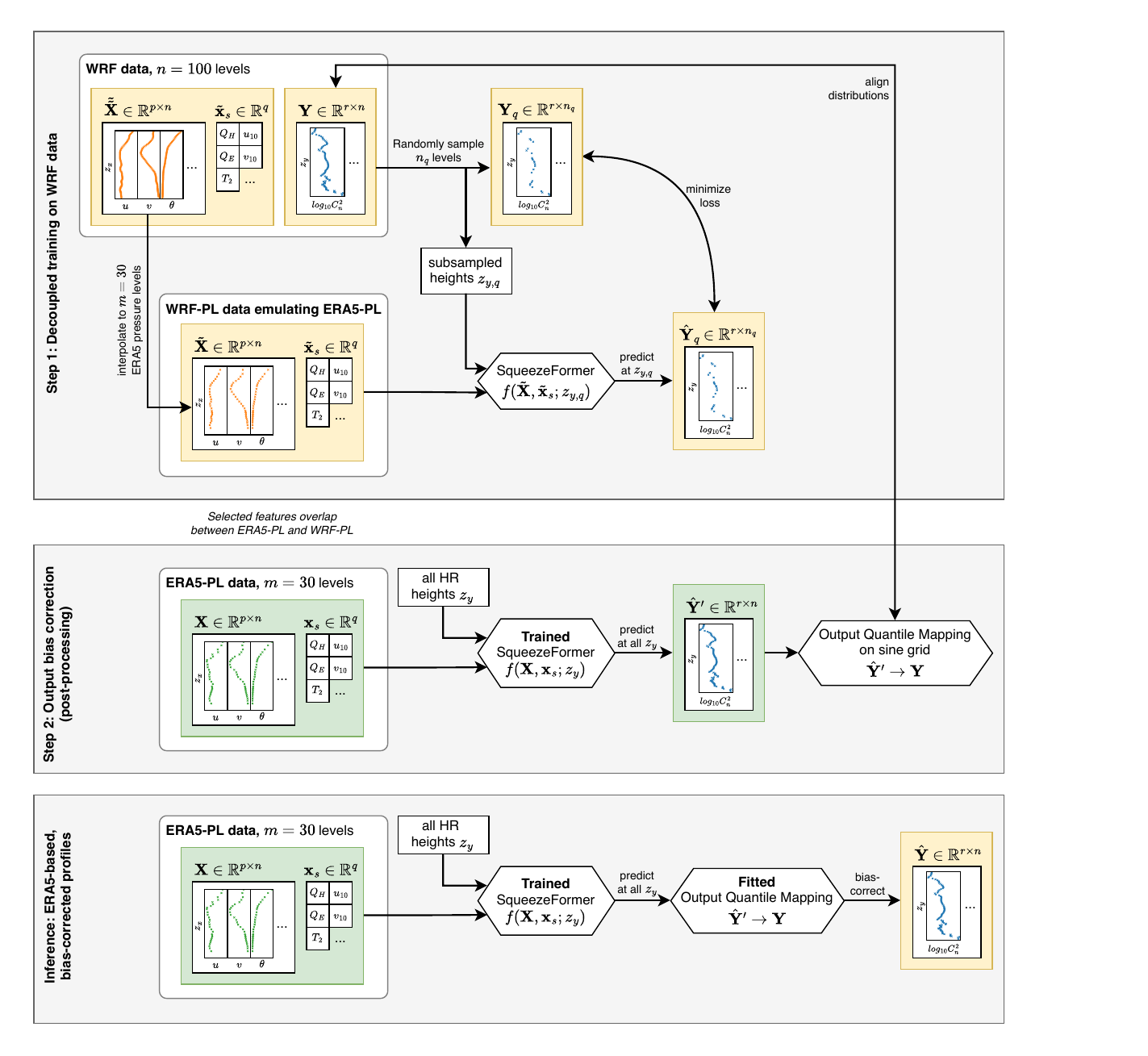}
   \caption{Two-step fitting pipeline of OTProf and ERA5-based inference step. Squeezeformer is trained on WRF data only (step 1). ERA5 data is used for inference, with output bias correction using quantile mapping.}\label{fig:sqf_pipeline}
\end{figure*}

Estimating fine, high-resolution (HR) vertical $C_n^2$ profiles from coarse, low-resolution (LR) meteorological profiles is a regression and super-resolution task.
For a single time and location, we consider a set of $p$ profile variables (features) $\mathbf{X}\in\mathbb{R}^{p\times m}$ given at $m$ levels at heights $\mathbf{z}_x \in \mathbb{R}^m$ above ground from which we aim to estimate $r$ profile variables (targets) $\mathbf{Y}\in\mathbb{R}^{r\times n}$ at $n$ levels at heights $\mathbf{z}_y \in \mathbb{R}^n$ above ground where $m<n$.
The input $\mathbf{X}$ contains meteorological profiles, such as temperature and wind speed, while the target profiles $\mathbf{Y}$ contain $C_n^2(z)$, which is not part of the inputs $\mathbf{X}$.
Because OT near the surface is primarily modulated by surface parameters such as surface heat flux and momentum flux, we also consider a vector of $q$ surface variables, $\mathbf{x}_s \in \mathbb{R}^q$, for a total of $p+q$ input variables.
The prediction task can then be formulated as
\begin{equation}
   f(\mathbf{X}, \mathbf{x}_s; \mathbf{z}_y) = \hat{\mathbf{Y}} \approx \mathbf{Y}
\end{equation}
at every time and location.
Here, the spatio-temporal dependency is implicit as $f$ does not aim to model a temporal evolution of $\mathbf{Y}$.

In practice, $\mathbf{X}$ and $\mathbf{Y}$ are typically produced by different models.
For example, $\mathbf{X}$ may be obtained from a global reanalysis, such as ERA5, while $\mathbf{Y}$ comes from high-resolution regional simulations, such as WRF.
This leads to spatio-temporal misalignment between datasets, which we address through a decoupled training approach combined with quantile mapping \citep{cannon2015}.
This decoupled pipeline is detailed in sec.~\ref{sec:decoupled_training}.
The Squeezeformer \citep{kim2022} DL architecture utilized in this study to obtain $f$ is described in sec.~\ref{sec:sqf}.
We introduce the two baseline models used to benchmark the Squeezeformer, including the Hufnagel-Valley model as a lower baseline, in sec.~\ref{sec:baselines}.
Finally, sec.~\ref{sec:model_eval} summarizes the evaluation metrics and strategy used to assess model performance.

\subsection{Decoupled training with output bias correction}\label{sec:decoupled_training}

Training machine learning models on datasets generated by different numerical models at varying resolutions is challenging because of spatiotemporal misalignment.
Regional models, such as WRF, drift over time relative to their forcing data because forcing is applied only at the boundaries, while the model evolves independently inside the domain.
Additionally, differences in physics parameterizations and resolution across models create structural differences.
For example, the wind field around complex terrain can differ significantly between a coarse global model and a high-resolution regional simulation.
An example is given in fig.~\ref{fig:res_diff}, where the 10\,m wind field around islands along the Dutch coast is presented.
In the HR wind field (panel b), a low-velocity wake is visible in the lee of the islands and the coast, together with terrain-induced changes of wind direction indicated by the white streamlines.
While the LR wind field (panel a) captures the general pattern of wind magnitude and direction, it lacks these fine-scale features.
Such misalignments are inherent and cannot be fully avoided, but we aim to partially mitigate them through decoupled training and output bias correction, as detailed below.

Our proposed framework to disentangle regression and super-resolution from these misalignment challenges is shown in fig.~\ref{fig:sqf_pipeline}.
Step 1 is the decoupled training.
Instead of directly training the Squeezeformer on ERA5 data as input and WRF data as the target, we train the model only on WRF data.
Specifically, we vertically coarsen the high-resolution WRF dataset to an LR dataset that emulates ERA5 in vertical resolution and variable selection.
These ERA5 proxy datasets are denoted with a tilde in fig.~\ref{fig:sqf_pipeline}.
Coarsening is achieved by vertically interpolating HR WRF profiles onto the LR ERA5 pressure levels in log space.
During training, the model learns to map LR WRF inputs to native HR WRF targets, learning regression and super-resolution on a consistent dataset.

To avoid the model memorizing the location-specific vertical output grid, the target profiles $\textbf{Y}$ are randomly subsampled in each batch and epoch.
Instead of always training on all $n$ HR levels with corresponding heights $z_{y}$, values at $n_q<n$ query levels are randomly drawn.
The model $f$ takes the heights of the subsampled levels $z_{y,q}$ as input to learn a robust mapping between $(\textbf{X}, \textbf{x}_s)$  and the subsampled profiles $\textbf{Y}_q$.
This architecture also enables OTProf models to be trained on multiple datasets with different grids, which we envision supporting fine-tuning on observations or data from different domains.
We recommend future work to explore this angle further.

For inference, we provide the real ERA5-PL data to the trained model to obtain HR profile estimates consistent with ERA5.
However, vertically interpolating WRF data to PL levels is not a perfect proxy for ERA5-PL.
Because ERA5-PL is horizontally smoother, it also yields less sharp vertical features (e.g., weaker gradients), resulting in systematically weaker turbulence.
We propose to partially compensate for this bias by post-processing the model outputs with quantile mapping \citep{cannon2015} in step 2 of the OTProf pipeline.
An ablation study demonstrating the effect of output quantile mapping (oQM) is presented in Appendix~A.
oQM statistically aligns the ERA5-based HR estimates $\hat{\textbf{Y}}'$ to the WRF-based HR reference $\textbf{Y}$ by matching their quantiles,
\begin{equation}
   \text{oQM}(\hat{\textbf{Y}}') = F_\textrm{WRF}^{-1}(F_{\textrm{ERA5}}(\hat{\textbf{Y}}')),
\end{equation}
where $F_\textrm{ERA5}$ and $F_\textrm{WRF}$ are the empirical cumulative distribution functions (CDF) of ERA5 and WRF of the target variables, respectively, and $F^{-1}$ denotes the inverse CDF.
Both CDFs are computed only on training data to avoid data leakage, and interpolation between CDFs is done on a sine-spaced grid with high point density at the edges to capture the long tails of $C_n^2$.
During inference on production data (final row in fig.~\ref{fig:sqf_pipeline}), the trained model $f$ produces HR profile estimates from ERA5-PL data, which are bias-corrected with the fitted oQM, resulting in the final estimates $\hat{\textbf{Y}}$.
  
\subsection{Squeezeformer}\label{sec:sqf}
\begin{figure*}[t]
   \includegraphics[width=\textwidth,trim=7mm 1mm 7mm 4mm,clip]{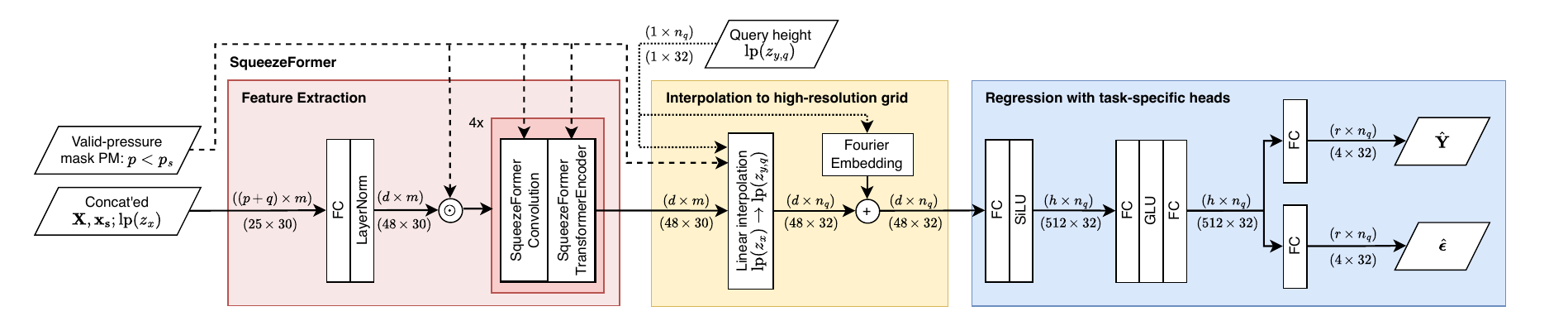}
   \caption{Squeezeformer architecture with linear interpolation in latent space for super-resolution. The Squeezeformer convolution and transformer blocks are abbreviated here and presented in detail in fig.~\ref{fig:sqf_arch_detail}. Dashed lines indicate where the valid-pressure mask (PM) enters, and dotted lines indicate the query heights. Numbers in brackets give the tensor shapes during training with $n_q=32$.}\label{fig:sqf_arch}
\end{figure*}

The Squeezeformer architecture, used to map LR to HR space, was originally proposed by \citet{kim2022} and has been successfully used in multiple Kaggle machine learning competitions \citep{christofhenkel2023,hoyso482024,greysnow2024b}. 
The core concept of the Squeezeformer combines the local feature extraction of convolutional neural networks with the long-range dependency modeling of transformer architectures \citep{kim2022}.
We consider this combination particularly well-suited to modeling turbulence in the atmospheric column because turbulence is modulated by both local gradients and larger-scale atmospheric structures \citep{stull1988}.
For the present study, we use the Squeezeformer variant of \citet{greysnow2024b}, which modifies the architecture from previous competitions for regressing atmospheric variables.
We extend the architecture for super-resolution by interpolating in feature space, as detailed below.

A schematic of the architecture is presented in fig.~\ref{fig:sqf_arch}, visualizing how LR input profiles are processed to yield HR outputs with corresponding confidence estimates.
Following \citet{greysnow2024b}, we incorporate the $q$ surface variables $\mathbf{x}_s$ by expanding them along the vertical dimension into profiles of constant values before concatenating them to the meteorological profiles $\mathbf{X}$, resulting in a tensor of shape $\mathbb{R}^{(p+q)\times m}$.
As ERA5's pressure levels are globally fixed, the lowest ERA5 levels can be locally below the surface, rendering their associated data meaningless.
This happens in high-altitude regions or complex terrain, where the surface pressure $p_s$ is lower than the pressure levels.
For reference, the four lowest PLs at \{1000, 925, 850, 700\}\,hPa correspond to heights of $\{\sim 100, \sim 750, \sim 1500, \sim 3000\}$\,m above mean sea level in the standard atmosphere, which are below the local terrain height in many places around the globe.
Therefore, we provide a valid-pressure mask (PM) as auxiliary input, which is 1 whenever a level is above the surface, i.e., when $p < p_s$, and 0 otherwise.

The Squeezeformer architecture follows a standard encoder-processor-decoder design, with a feature extraction backbone (encoder), latent-space interpolation to query heights (processor), and regression to the turbulence output variables (decoder).
In the encoder (red block), the concatenated inputs are first embedded into a higher-dimensional space $\mathbb{R}^{d\times m}$ with $d>(p+q)$ using a fully connected layer (FC) with layer normalization.
We apply the PM to the output to cancel below-ground levels before the embeddings are processed by a sequence of Squeezeformer blocks composed of convolution and transformer blocks.
A detailed description of the Squeezeformer blocks is given in Appendix~B but is skipped here for brevity.
The PM is also passed to the convolutions and multi-head attention layers in the Squeezeformer blocks to always cancel below-ground signals.

The extracted features are linearly interpolated (yellow block) to the $n_q$ query heights in latent space on a scaled vertical coordinate, shared between ERA5 and WRF:
\begin{equation}
   \textrm{lp}(z) = \ln\left(1+\frac{z}{10\,\textrm{m}}\right) \bigg/ \ln\left(1+\frac{30000\,\textrm{m}}{10\,\textrm{m}}\right).
   \label{eq:lz}
\end{equation}
The lp function transforms the height above ground $z$ into logarithmic space and scales it such that it is approximately 0 at the surface and 1 at the top of the atmosphere.
The scaling bounds are fixed to keep the transformation dataset-independent.
The PM ensures again that only above-ground levels are used for interpolation,
and a Fourier embedding of the query heights $\textrm{lp}(z_{y,q})$ adds a vertical positional encoding to the interpolated latent variables.
The Fourier embedding details are given in Appendix~B.
This interpolation setup makes the trained model technically output-grid agnostic, meaning that it can interpolate the latent features to arbitrary output heights.
That enables future training or fine-tuning on multiple datasets with different grids.
In the present scenario, where the model is trained on a single dataset with little variation in the vertical coordinate, performance is expected to be best near the heights of the original data.

Finally, the decoder (blue block) projects the interpolated embeddings into a high-dimensional regression space $\mathbb{R}^{h\times n_q}$ through several FC layers with sigmoid linear unit (SiLU) \citep{ramachandran2017} and gated linear unit (GLU) activations \citep{dauphin2017}.
Gated activation enables the model to suppress irrelevant information and focus on important features.
Two parallel task-specific regression heads (FC layers) bring the dimensionality of $h$ down to the $r$ target variables.
The output heads yield the target profile values ($\mathbf{Y}$) and corresponding confidence estimates ($\hat{\boldsymbol{\epsilon}}$), i.e., estimates of the squared prediction error (see below).
The Squeezeformer is trained to minimize a two-part loss function comprising a regression loss and a confidence loss that serves as regularization \citep{greysnow2024b}.
Both loss components are based on the root-mean-squared error (RMSE), where the regression loss minimizes the squared error between predicted and reference profiles, $\boldsymbol{\epsilon}_i = \left(\mathbf{Y}_i - \hat{\mathbf{Y}}_i\right)^2 \in \mathbb{R}^{r\times n}$:
\begin{equation}
   L_{rgr} = \sqrt{\langle \boldsymbol{\epsilon}_i \rangle}.
\end{equation}
Here, the index $i$ indicates a single sample of target profiles and $\langle \square \rangle$ denotes the mean over all target variables, levels, and samples.
The task of the confidence head is to estimate the squared error $\boldsymbol{\epsilon}_i$ as a confidence estimate, denoted $\hat{\boldsymbol{\epsilon}}_i$,
so the confidence loss is defined as the RMSE between $\boldsymbol{\epsilon}_i$ and $\hat{\boldsymbol{\epsilon}}_i$:
\begin{equation}
   L_{conf} = \sqrt{\langle \left(\boldsymbol{\epsilon}_i - \hat{\boldsymbol{\epsilon}}_i\right)^2 \rangle}.
\end{equation}
As $\hat{\boldsymbol{\epsilon}}_i$ estimates the squared error, it carries units of the target variable squared.
Thus, a confidence estimate in the units of the target variable is $\sqrt{\hat{\boldsymbol{\epsilon}}_i}$.
The total loss is the sum of regression and confidence loss, $L = L_{rgr} + L_{conf}$, and is minimized using the AdamW \citep{loshchilov2019} optimizer.
Cosine annealing with warm restarts \citep{loshchilov2017} is used to schedule the learning rate during training, and manual early stopping based on the validation loss is applied to prevent overfitting. 
We train the models with a learning rate of $10^{-3}$, weight decay of $10^{-2}$, and a batch size of 23552 for at most 600 epochs, with a restart period of 10 epochs.

\subsection{Baseline models}\label{sec:baselines}
To benchmark Squeezeformer performance, we consider two baseline models that represent the lower and upper performance bounds.
The upper baseline is a Squeezeformer trained and evaluated only on HR data, representing the architecture's best-case performance without super-resolution.
For this baseline, the input variables (profiles and surface) are the same as for the main model, so the regression task to estimate HR $C_n^2$ profiles from standard meteorological variables is identical.

As a lower baseline, we consider the Hufnagel-Valley (HV) model \citep{hufnagel1974,valley1980,ulrich1988} combined with the \citet{wyngaard1971} parameterization for surface $C_n^2$. 
The HV model is a simple analytical, empirical, and parametric model widely used in the optical turbulence community (e.g., \citet{andrews2009,dimitrov2016,camboulives2018,osborn2021,walsh2023}).
Following \citet{beland1993}, the HV model is given as
\begin{equation}
   \begin{aligned}
      C_n^2 \left( \tilde{z} \right) = {} & c_1 W^2 \tilde{z}^{10} \exp\left(-\tilde{z}\right) + c_2 \exp \left(-\frac{\tilde{z}}{1.5}\right) \\
      & + \exp\left(-\frac{\tilde{z}}{0.1}\right) \left(\left.C_n^2\right|_{z=0} - c_2\right)
   \end{aligned}
   \label{eq:hv}
\end{equation}
where $\tilde{z}$ is the height above ground in km, and $c_1$ and $c_2$ are constants with $c_1=8.2\times10^{-26}$\,m$^{-2/3}\,$km$^{-10}\,$s$^{2}\,$m$^{-2}$ and $c_2=2.7 \times 10^{-16}$\,m$^{-2/3}$, respectively.
The parameter W represents the root-mean-square (RMS) wind speed in m\,s$^{-1}$ between 5\,km and 20\,km height.
Considering a profile of absolute wind speed $m(\tilde{z})$, W is computed as
\begin{equation}
   W = \left(\frac{1}{15}\int_5^{20} m(\tilde{z})^2 \, d\tilde{z}\right)^{1/2}.
\end{equation}

Equation~\eqref{eq:hv} also requires a surface estimate of $C_n^2$ which we obtain following the flux-based approach of \citet{wyngaard1971}:
\begin{equation}
   \left.C_n^2\right|_{z=0}= {\left(A\, \frac{p_s}{T_2^2}\right)}^2 T_*^2 z^{-2/3} g(\zeta).
   \label{eq:cn2_w71}
\end{equation}
The temperature scale $T_*=-\overline{w'\theta'}/u_*$ is computed using the friction velocity $u_*$ and the kinematic sensible heat flux $\overline{w'\theta'}$ at the surface, while $p_s$ is the surface pressure in hPa, $T_2$ is the temperature at 2\,m height in K, and $z$ is the measurement height in m.
$A \approx 7.9\times10^{-5}\, \text{K}\, \text{hPa}^{-1}$ is almost constant for optical wavelengths \citep{andrews2005} and $g(\zeta)$ is a stability-dependent similarity function with $\zeta = z/L$ and Obukhov length $L$.
We call this combination HV+W71 for the rest of the manuscript.

All meteorological variables required by the HV model are typically available from atmospheric datasets.
Except for the wind profile to compute $W$, only surface variables are needed.
To fairly compare the Squeezeformer against HV+W71, we utilize HV+W71 in a ``superresolution mode'' where $W$ is computed from LR profiles of $m(\tilde{z})$ but eq.~\eqref{eq:hv} is evaluated at the $\tilde{z}$ positions of the HR levels.

\subsection{Model evaluation}\label{sec:model_eval}
The performance of the fitted models is assessed with respect to their ability to capture the HR reference $C_n^2$ profiles and two integrated astroclimatic parameters derived from these profiles, the Fried parameter $r_0$ and the scintillation index $\sigma_I^2$.
We use both parameters for their practical relevance to the OT community \citep{beland1993,hardy1998} and because they weight different parts of the $C_n^2$ profiles differently.
This weighting lets us assess how well different parts of the profiles are captured.
To quantify agreement between predictions and reference data, we use five metrics: bias, centered root-mean-square error (cRMSE), spatial and temporal Pearson correlation coefficient ($r$), and differences in skewness and kurtosis between the estimated distributions and the reference.
Additionally, a structure-function analysis is performed to assess how well the models capture the vertical variability of the $C_n^2$ profiles across different vertical scales.
All parameters, metrics, and the structure-function analysis are detailed below.

\subsubsection{Integrated astroclimate parameters}
\paragraph*{Fried Parameter}
The Fried parameter \citep{fried1966} (in centimeters) measures the strength of wavefront distortions caused by OT.
If the ratio of receiver aperture diameter to the Fried parameter, $D/r_0$, is less than one, the optical system operates close to its theoretical optimum, i.e., it is diffraction-limited. 
At the same time, turbulence-induced distortions degrade the performance for $D/r_0 > 1$.
Assuming a vertical profile of $C_n^2$, $C_n^2(z)$, the Fried parameter is given as \citep{andrews2005}
\begin{equation}
  r_0 = \left(0.423 k^2 \int_0^H C_n^2(z) dz \right)^{-3/5},
  \label{eq:fried_vertical}
\end{equation}
where $k = 2\pi/\lambda$ is the wavenumber and $H$ is the propagation distance.

\paragraph*{Scintillation Index}
The scintillation index, $\sigma_I^2$, expresses the normalized variance of received optical intensity fluctuations \citep{andrews2005}
\begin{equation}
  \sigma_I^2=\frac{\langle I^2 \rangle - {\langle I \rangle}^2}{{\langle I \rangle}^2},
\end{equation}
where $I(t)$ is the received intensity/irradiance signal and $\langle \square \rangle$ denotes the ensemble average over a given time period.

A theoretical connection between $\sigma_I^2$ and the $C_n^2(z)$ profile can be made by assuming a Kolmogorov turbulence spectrum and Rytov theory.
We first define the Rytov variance for plane waves as \citep{andrews2005}
\begin{equation}
   \sigma_R^2 = 2.25 k^{7/6} \int_0^H C_n^2(z) z^{5/6} dz,
   \label{eq:rytov_vertical}
\end{equation}
where $k$ and $H$ again correspond to the laser wavenumber and propagation distance, and the receiver is located at $z=0$ (ground level), corresponding to a downlink scenario.
To account for the saturation of scintillation in strong turbulence conditions, i.e., the decoupling of $C_n^2$ and $\sigma_I^2$ in strong turbulence, we relate $\sigma_R^2$ to $\sigma_I^2$ as \citep{andrews2005}
\begin{equation}
   \begin{aligned}
      \sigma_I^2 = \exp\Biggl(&\frac{0.49\, \sigma_R^2}{\left[1 + 1.11\, \left({\sigma_R^2}\right)^{6/5}\right]^{7/6}} \\
      &+ \frac{0.51\, \sigma_R^2}{\left[1 + 0.69\, \left({\sigma_R^2}\right)^{6/5}\right]^{5/6}}\Biggr) - 1.
   \end{aligned}
\end{equation}
This expression also includes the weak-turbulence case ($\sigma_R^2 < 1$), where $\sigma_I^2 \approx \sigma_R^2$.

Comparing the integrands of $r_0$ and $\sigma_R^2$ shows that the integration weights different parts of the $C_n^2(z)$ profile differently.
Compared to $r_0$, the expression for $\sigma_R^2$ adds a $z^{5/6}$ term to the integrand, weighting high-altitude turbulence more strongly than the uniform weighting of $r_0$.
Because $C_n^2$ is typically strongest near the surface, $r_0$ primarily reflects near-surface turbulence, whereas $\sigma_I^2$ is more sensitive to upper-air turbulence.
These different weightings let us assess how well different parts of the $C_n^2(z)$ profiles are captured, with a practical interpretation attached. 

\subsubsection{Performance metrics}
Five metrics are used to quantify the performance of the different models in estimating HR $C_n^2$ profiles, $r_0$, and $\sigma_I^2$: 
bias, centered root-mean-square error (cRMSE), vertical and temporal Pearson correlation coefficient ($r$), and the differences in skewness ($\Delta \text{skew}$) and kurtosis ($\Delta \text{kurt}$) of the distributions.
Following their approximately log-normal distributions, $C_n^2$ and $\sigma_I^2$ are evaluated as $\log_{10} C_n^2$ and $\log_{10} \sigma_I^2$, whereas $r_0$ is evaluated in linear space (in cm).

As our dataset is multi-dimensional, and the interpretation of a metric changes depending on the axis along which it is computed, we first define the following notation for clarity.
For a single target variable, the reference value at location $l$, time $t$, and level $i = 1, \dots, n$ is denoted as $y_{l,t,i}$, and the corresponding estimate as $\hat{y}_{l,t,i}$.
For brevity, we omit the indices and write $y$ and $\hat{y}$ where the meaning is clear.
The arithmetic mean over one or more indices is denoted by angle brackets with these indices as subscripts, while the remaining indices are kept.
For example, $\langle y \rangle_t$ is the temporal mean profile at each location (local climatology), and $\langle y \rangle_{l,t,i}$ is the global mean.
For the integrated parameters $r_0$ and $\sigma_I^2$, the level index $i$ is dropped.

\paragraph*{Bias and error} Using this notation, we define the global bias between estimated profiles $\hat{y}$ and their reference values $y$ as
\begin{equation}
   \textrm{Bias} = \langle \hat{y} \rangle_{l,t,i} - \langle y \rangle_{l,t,i}.
\end{equation}
For $\log_{10} C_n^2$, a positive (negative) bias means the estimated turbulence is too strong (weak) relative to the reference.
To quantify the amplitude of the model error, we use the centered root-mean-squared error (cRMSE), the RMSE after removing temporal bias at each location and level.
In the present study, the cRMSE is used to quantify temporal error.
Therefore, the local climatology is first subtracted from both estimate and reference: $\hat{y}' = \hat{y} - \langle \hat{y} \rangle_t$ and $y' = y - \langle y \rangle_t$.
The cRMSE is then computed per level and location and averaged as
\begin{equation}
   \textrm{cRMSE} = \left\langle \sqrt{\left\langle {\left(\hat{y}' - y'\right)}^2 \right\rangle_t} \right\rangle_{l,i}.
\end{equation}

\paragraph*{Correlation} The Pearson correlation coefficient $r$ quantifies how well features, such as turbulence peaks, are located in time and height.
In general, correlation along index $\square$ between $y$ and $\hat{y}$ is defined as
\begin{equation}
   r_\square(y, \hat{y}) = \frac{\left\langle \left(y - \langle y \rangle_\square\right)\left(\hat{y} - \langle \hat{y} \rangle_\square\right) \right\rangle_\square}{\sqrt{\left\langle {\left(y - \langle y \rangle_\square\right)}^2 \right\rangle_\square \left\langle {\left(\hat{y} - \langle \hat{y} \rangle_\square\right)}^2 \right\rangle_\square}}.
\end{equation}
To quantify how well OTProf places turbulent features in time, the correlation in time at each level and location is computed as $r_t(y, \hat{y})$ and further averaged to obtain a global metric:
\begin{equation}
   r_{\textrm{temp}} = \left\langle r_t(y, \hat{y}) \right\rangle_{l,i}.
\end{equation}
To quantify the accuracy of feature placement in the vertical direction, we compute $r$ for each column, at each location and time.
To avoid overly optimistic correlation scores, when the estimated profiles capture the dominant but simple trend of exponential decay with height, we first subtract the local climatology as for the cRMSE.
A global score is obtained by averaging the vertical correlation coefficients across all locations and times:
\begin{equation}
   r_{\textrm{vert}} = \left\langle r_i(y', \hat{y}') \right\rangle_{l,t}.
\end{equation}
Values of $r_\textrm{temp}$ and $r_\textrm{vert}$ closer to 1 indicate that turbulent features are, on average, more accurately placed in time and height, respectively.
Both coefficients are insensitive to bias and to errors in the amplitude of the features, which are instead captured by the bias and the cRMSE.

\paragraph*{Distribution shape} The agreement of the shape of the estimated distribution with that of the reference distribution of a profile variable is quantified by using the differences in skewness and excess kurtosis.
Both metrics are based on standardized statistical moments that are zero for a Gaussian and describe the tails of a distribution, which is crucial for judging how well extreme (weak or strong) turbulence conditions are estimated.
We evaluate the metrics globally, i.e., across the entire dataset, to quantify how well the overall climatology is estimated.
For $\log_{10} C_n^2$, this global distribution also contains the decrease of turbulence strength with height, so its tails partly reflect differences between near-surface and upper-air conditions.

Skewness quantifies asymmetry, where positive (negative) skewness means the distribution has a longer right (left) tail, so extreme values occur mainly above (below) the bulk of the distribution, compared to a symmetric Gaussian with zero skewness and equal tails.
The larger the absolute value of the skewness, the more pronounced this one-sided tail is relative to the other.
The excess kurtosis quantifies the overall weight of the tails, with positive (negative) excess kurtosis corresponding to heavier (lighter) tails with more (fewer) extreme values in either direction than a Gaussian with zero excess kurtosis would have.
For strongly skewed distributions, e.g., that of $r_0$, the excess kurtosis is mostly driven by the long tail.

Skewness and excess kurtosis are defined here based on the $p$-th global central moments $\mu_p$ of the dataset as
\begin{equation}
   \textrm{skew}(y) = \frac{\mu_3}{\mu_2^{3/2}}, \quad \textrm{kurt}(y) = \frac{\mu_4}{\mu_2^2} -3,
\end{equation}
with
\begin{equation}
   \mu_p(y) = \left\langle {\left(y - \langle y \rangle_{l,t,i}\right)}^p \right\rangle_{l,t,i}.
\end{equation}
The differences between both metrics for an estimated set of profiles $\hat{y}$ and a reference set $y$ are defined as
\begin{equation}
   \Delta \textrm{skew} = \textrm{skew}(\hat{y}) - \textrm{skew}(y), \quad
   \Delta \textrm{kurt} = \textrm{kurt}(\hat{y}) - \textrm{kurt}(y).
\end{equation}
For $\log_{10} C_n^2$ and $\log_{10} \sigma_I^2$, $\Delta \textrm{skew} > 0$ means that the estimated distribution has a longer tail toward strong turbulence than the reference, i.e., unusually strong turbulence is overrepresented or unusually weak turbulence is underrepresented.
For $\Delta \textrm{skew} < 0$, the estimate instead has a longer tail toward weak turbulence.
Since $r_0$ decreases with increasing turbulence strength, the interpretation of $r_0$ is reversed.
A positive $\Delta \textrm{kurt}$ means that the estimate contains too many extreme turbulence conditions, i.e., conditions that are unusually strong or weak compared to typical conditions, whereas a negative $\Delta \textrm{kurt}$ means that such extremes are too rare.
The skewness indicates whether these extremes are mainly strong or weak.
 
All metrics except $r_\textrm{vert}$ are also used to quantify the accuracy of the estimated integrated parameters $r_0$ and $\sigma_I^2$.

\subsubsection{Structure function analysis}
Structure function (SF) analysis assesses how well models capture the vertical variability of $C_n^2$ profiles across different vertical scales \citep{lovejoy2012}.
The structure function of a signal relates to its power spectrum via the Fourier transform \citep{frisch1995,lovejoy2012}, so one can loosely think of SF analysis as spectral analysis in physical space rather than frequency space.
SFs have the advantage of being easier to compute from non-uniformly spaced data, as is typical in atmospheric profiles.
Additionally, interpreting SFs in physical space is often more intuitive than interpreting power spectra in frequency space.
It should be emphasized that SFs are employed in this study solely as a diagnostic tool to assess vertical variability across different scales.
We do not aim to link the SFs to turbulence theory, where certain scaling laws are expected, or additional interpretation constraints apply.

The 2nd order structure function of an ensemble of vertical profiles $y(z)$ is given as
\begin{equation}
   S_2(\Delta z) = \langle \left(y(z + \Delta z) - y(z)\right)^2 \rangle,
\end{equation}
where $S_2(\Delta z)$ is the average squared increment between two points in the profile separated by a vertical distance $\Delta z$.
Often, SFs follow power laws of the form $S_2(\Delta z) = c\, \Delta z^\zeta$ for parts of the $\Delta z$ range, where $\zeta$ is called the scaling exponent.
Comparing the $\zeta$ and $c$ of the SFs of the predicted and reference profiles lets us assess, for example, whether the profiles differ in smoothness or whether vertical variability (i.e., features) is missing at certain scales.
Therefore, SF analysis is a powerful diagnostic tool for assessing prediction quality across scales.

In practice, we compute the vertical SF of $\log_{10} C_n^2$ for all profiles in the dataset, using the notation introduced above.
Because meteorological datasets often use constant-pressure levels rather than constant height $z$ in the vertical direction, the $z$ spacing between levels is typically non-uniform and time-dependent.
Therefore, each value $y_{l,t,i}$ has an associated height above ground $z_{l,t,i}$.
For each profile, i.e., at each location $l$ and time $t$, we compute the squared increments $\left(y_{l,t,i} - y_{l,t,j}\right)^2$ and the corresponding vertical distances $\Delta z = |z_{l,t,i} - z_{l,t,j}|$ for all pairs of levels $i < j$.
The SF is then obtained by binning the squared increments across all locations and times according to their $\Delta z$ and averaging them within each $\Delta z$ bin.

With this methodological foundation in place, the following section describes the datasets used for training and inference, including details on the WRF simulations that provide the HR training targets and the ERA5 reanalysis that provides the LR input data.

\section{Datasets}\label{sec:datasets}

\begin{figure}
   \begin{minipage}[t]{.59\columnwidth}
      \begin{subfigure}[t]{\textwidth}
         \caption{ERA5 \horres{0.25}{$^\circ$}}
         \includegraphics[width=\textwidth,trim=0mm 0mm 109mm 0mm,clip]{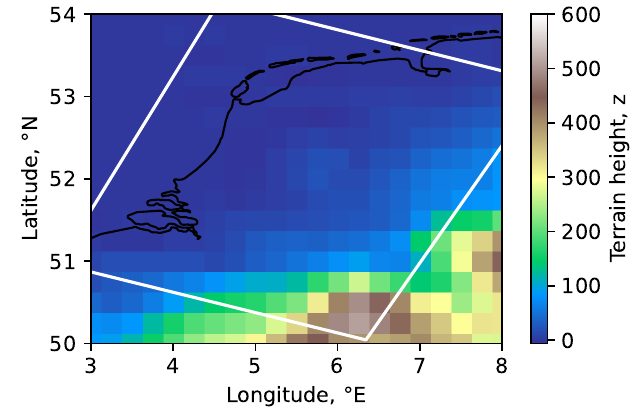}
      \end{subfigure}

      \begin{subfigure}[t]{\textwidth}
         \caption{WRF \horres{2}{km}}
         \includegraphics[width=\textwidth,trim=109mm 0mm 0mm 0mm,clip]{figures/wrf_era5_terrain.pdf}
      \end{subfigure}
   \end{minipage}%
   \begin{minipage}[t]{.41\columnwidth}
      \begin{subfigure}[t]{\textwidth}
         \caption{Average height of vertical levels in WRF and ERA5.}\label{fig:levels}
         \includegraphics[width=\textwidth]{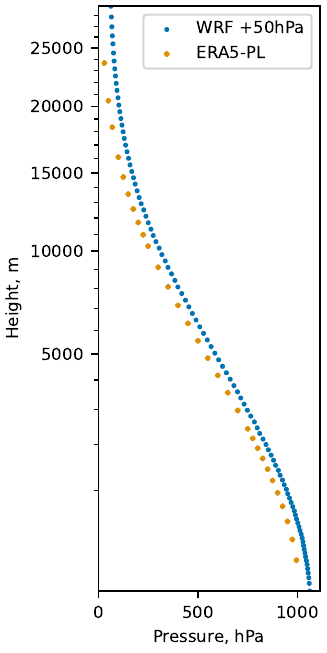}
      \end{subfigure}%
   \end{minipage}
   \caption{Horizontal (a and b) and vertical (c) extent of ERA5 and WRF domains used in this work. The WRF points in (c) are artificially shifted by 50\,hPa for visualization.}%
   \label{fig:era5_wrf}
\end{figure}

The two main datasets used in this work are a high-resolution (\horres{2}{km}) mesoscale dataset generated using the Weather Research and Forecasting (WRF) model \citep{skamarock2021} and the lower-resolution (\horres{0.25}{$^\circ$} corresponding to ca. \horres{17}{km} at 52$^\circ$N in the Netherlands) ERA5 reanalysis dataset \citep{era5}.
The WRF data have higher horizontal and vertical resolution, with 100 vertical levels between the surface and ca.~20\,km, compared to 37 pressure levels in ERA5 up to ca.~30\,km.
The extent of the domains in horizontal and vertical directions is illustrated in fig.~\ref{fig:era5_wrf}.
Both datasets cover the same geographic region, here, the Netherlands, and the same time periods.
Details about the generation of the WRF dataset are presented in sec.~\ref{sec:wrf}, while ERA5 is introduced in sec.~\ref{sec:era5}.
Section~\ref{sec:dl_data} summarizes the construction of the datasets used for training and inference of the Squeezeformer model.

\subsection{WRF: year-long $C_n^2$ database for NL}\label{sec:wrf}
The Weather Research and Forecasting (WRF) model \citep{skamarock2021} is used to generate a statistically representative\footnote{We refer to temporal representativeness for the Netherlands here. Geographical representativeness requires more extensive WRF simulations.}, high-resolution training dataset of $C_n^2$ over the Netherlands.
The WRF configuration is based on \citet{pierzyna2024}, where $C_n^2$ is estimated following the variance-based parameterization of \citet{he2015}.

We aim to simulate a full year of hourly output at \horres{2}{km} horizontal resolution and 100 vertical levels, reaching up to ca.~20\,km in height.
To keep this task manageable in terms of computational costs and storage requirements, we employ two tricks.
First, we use the Copernicus European Regional ReAnalysis (CERRA, \citet{cerra}) dataset, which has higher horizontal resolution than the global ERA5, to force WRF.
Simulations forced with ERA5 typically require computing and storing 3 nested domains, whereas using CERRA (\horres{5.5}{km}) lets us reach the target resolution of \horres{2}{km} with a single domain \citep{baki2025a}.
We assume the 2\,km resolution is still sufficient to capture the relevant processes that modulate optical turbulence.

The second trick is not to simulate a continuous year-long trajectory, but instead to run a series of shorter simulations staggered over multiple years to increase the statistical representativeness of meteorological conditions.
In particular, we run 73 5.5-day simulations with 12 hours of warmup each, yielding 365 days net.
The simulations are staggered in a round-robin fashion over four years (2017, 2018, 2019, 2020) to avoid sampling a particularly hot, cold, wet, or dry year.
Consequently, simulations for the same year are always 15 days apart, with 18 simulations per year (19 for 2017 to cover the full 365 days).
This short simulation length also limits model drift relative to the reanalysis forcing, as discussed in sec.~\ref{sec:decoupled_training}.
Based on our experience, the simulation length of 5.5 days is a trade-off between limiting the drift and the computational cost of the warmup.
Shorter simulations would drift less, but the 12\,h warmup would then make up a larger fraction of each simulation without yielding usable data.
Additionally, the single domain is forced directly by CERRA, which couples the simulation more strongly to the forcing data than a set of nested domains forced by ERA5.

The 2nd-order structure function parameter of the refractive index, $C_n^2$, is estimated by post-processing the WRF model output. 
The variance-based parameterization of \citet{he2015} first yields an estimate of $C_T^2$, the 2nd-order structure function parameter of \emph{temperature}, 
\begin{equation}
    C_T^2 = 3.2 \chi \epsilon^{-1/3},
    \label{eq:ct2_corrsin}
\end{equation}
where $\epsilon$ is the turbulent energy dissipation rate per unit mass and $\chi$ the destruction rate of potential temperature variance $\sigma_\theta^2$ given as \citep{mellor1973,mellor1982}
\begin{equation}
    \epsilon = \frac{(2e)^{3/2}}{B_1 L_M}, \quad
    \chi = \frac{(2e)^{1/2}}{B_2 L_M}\,\sigma_\theta^2.
    \label{eq:ct2_var_chi_eps}
\end{equation}
Here, $e$ is the turbulent kinetic energy (TKE) per unit mass, and $L_M$ is the master length scale that combines the characteristic scales of the surface, turbulence, and buoyancy.
The coefficients $B_1$ and $B_2$ were found to be equal to 24 and 15, respectively, from numerical simulations.
Following \citet{he2015}, we configure WRF to use the high-order Mellor-Yamada-Nakanishi-Niino (MYNN) 2.5 turbulence closure scheme \citep{nakanishi2006, nakanishi2009}, which yields $e$, $\sigma_\theta^2$, and $L_M$.
Finally, the $C_T^2$ estimates are converted to $C_n^2$ following the Gladstone relation 
\begin{equation}
   C_n^2 = {\left(A\, \frac{p}{T^2}\right)}^2 C_T^2
\end{equation}
with $A \approx 7.9\times10^{-5}\, \text{K}\, \text{hPa}^{-1}$, mean pressure $p$, and mean temperature $T$.
Different formulations of this relation exist, e.g., the non-hydrostatic formulation of \citet{cherubini2013}, which uses $T\theta$ instead of $T^2$.
For consistency with our previous work \citep{Basu2020,pierzyna2024}, we use the formulation given above.

The He--Basu parameterization is well established and has been evaluated against observations in previous studies, including comparisons of WRF-derived $C_n^2$ against thermosonde profiles \citep{Basu2020} and against near-surface measurements \citep{pierzyna2024}.
The broader class of TKE-based mesoscale $C_n^2$ parameterizations is used operationally for optical turbulence forecasting at astronomical sites \citep{bougeault1995,masciadri1999,masciadri2013,masciadri2020}. 
We therefore regard the resulting profiles as a suitable reference for training and evaluating OTProf. 
However, we acknowledge that uncertainties from the mesoscale simulations and the parameterization carry over into the estimated $C_n^2$, and future work should explicitly benchmark OTProf against observations.

The computational cost of each simulation of the 144 $\times$ 192 domain amounts to 10\,h on 32 cores of an AMD EPYC 7H12 CPU, resulting in 320 core hours per simulation or 23\,360 core hours for the full year. 
Post-processing each simulation to extract $C_n^2$ needs 0.8 core hours per simulation or 58.4 core hours for the full dataset.
As a result, producing the full $C_n^2$ dataset requires just below 23\,500 core hours.

\subsection{ERA5 reanalysis}\label{sec:era5}
With $C_n^2$ profiles from WRF forming the HR target data, we require collocated LR meteorological input profiles and surface features. 
We obtain these from the ERA5 reanalysis dataset \citep{era5}, available globally at a horizontal resolution of \horres{0.25}{$^\circ$}.
The vertical resolution of ERA5 depends on the variant used.
For this work, we use the ERA5 pressure-level dataset (ERA5-PL), which contains 37 pressure levels between 1000\,hPa and 1\,hPa and is commonly used and readily available.
Additionally, ERA5 provides a variety of surface variables at each grid point and has a temporal resolution of 1 hour, matching the WRF output frequency.
Data are selected to cover the same geographic region and time periods as the WRF simulations.
A list of the selected ERA5 variables is provided in tab.~\ref{tab:features}.

\subsection{Construction of the deep learning datasets}\label{sec:dl_data}
\begin{table*}
   \caption{Overview of $p+q=9+16=25$ features and $r=4$ targets contained in the datasets. Variables marked with (*) are derived from other variables. All variables are normalized to the range $[0, 1]$ using the minimum and maximum values specified in the normalization column, where the prefix P indicates percentiles.}%
   \label{tab:var_overview}
   \begin{subtable}[t]{\textwidth}
      \caption{Low-resolution meteorological input features from ERA5-PL or WRF-PL}\label{tab:features}
      \centering
      \begin{tabular}{llllc}
\toprule
Variable & Description & Unit & Type & Normalization \\
\midrule
$z$ & Height above ground & m & Profile & [min, max] \\
\revM{$\textrm{lp}(z)$} & Logarithmic scaled height above ground, cf.~eq.~\eqref{eq:lz} & - & Profile (*) & - \\
$u(z)$ & Zonal wind speed & m\,s$^{-1}$ & Profile & [P1, P99] \\
$v(z)$ & Meridional wind speed & m\,s$^{-1}$ & Profile & [P1, P99] \\
$\theta(z)$ & Potential temperature & K & Profile & [min, max] \\
$p(z)$ & Pressure & hPa & Profile & [min, max] \\
\revM{$\textrm{PM}(z)$} & Valid-pressure mask ($p < p_s$) & - & Profile (*) & - \\
$S(z)$ & Wind shear, cf.~eq.~\eqref{eq:shear} & s$^{-1}$ & Profile (*) & [0, P99] \\
$\Gamma(z)$ & Potential temperature gradient, cf.~eq.~\eqref{eq:gamma} & K\,m$^{-1}$ & Profile (*) & [P1, P99] \\
\revM{$\ln(p_s)$} & Logarithmic surface pressure & hPa & Surface (*) & [min, max] \\
\revM{$z_t$} & Terrain height & m & Surface (*) & [min, max] \\
$u_*$ & Friction velocity & m\,s$^{-1}$ & Surface & [0, P99] \\
$Q_H$ & Surface sensible heat flux & W\,m$^{-2}$ & Surface & [P1, P99] \\
$Q_E$ & Surface latent heat flux & W\,m$^{-2}$ & Surface & [P1, P99] \\
$[\log_{10} C_T^2]_s$ & $C_T^2$ at surface \citep{wyngaard1971} & - & Surface (*) & [-8, P99.5] \\
$h$ & Boundary layer height & m & Single Level & [0, P99] \\
$u_{10}$ & Zonal wind speed, 10\,m above ground & m\,s$^{-1}$ & Single Level & [P1, P99] \\
$v_{10}$ & Meridional wind speed, 10\,m above ground & m\,s$^{-1}$ & Single Level & [P1, P99] \\
$T_2$ & Air temperature, 2\,m above ground & K & Single Level & [P1, P99] \\
$p_0$ & Mean sea level pressure & hPa & Single Level & [min, max] \\
LSM & Land-sea mask & - & Single Level & - \\
$\cos(\text{hr}')$ & Cosine of normalized hour of the day & - & Single Level (*) & [min, max] \\
$\sin(\text{hr}')$ & Sine of normalized hour of the day & - & Single Level (*) & [min, max] \\
$\cos(\text{doy}')$ & Cosine of normalized day of the year & - & Single Level (*) & [min, max] \\
$\sin(\text{doy}')$ & Sine of normalized day of the year & - & Single Level (*) & [min, max] \\
\bottomrule
\end{tabular}

   \end{subtable}%
   \vspace{1em}

   \begin{subtable}[t]{\textwidth}
      \caption{High-resolution turbulence target variables from WRF}\label{tab:targets}
      \begin{tabular}{llllc}
\toprule
Variable & Description & Unit & Type & Normalization \\
\midrule
$\log_{10}\left[C_n^2(z)\right]$ & $C_n^2$ profile, acc.\ to eq.~\ref{eq:ct2_corrsin} & - & Profile (*) & [P0.5, P99.5] \\
$\log_{10}\left[2e(z)\right]$ & Twice the turbulent kinetic energy from WRF & - & Profile & [P0.5, P99.5] \\
$\log_{10}\left[L_m(z)\right]$ & Master length scale from WRF & - & Profile & [P0.5, P99.5] \\
$\log_{10}\left[\sigma_\theta^2(z)\right]$ & Potential temperature variance from WRF & - & Profile & [P0.5, P99.5] \\
\bottomrule
\end{tabular}

   \end{subtable}%
\end{table*}

The data from WRF and ERA5 are combined into a single dataset by collocating each ERA5 grid point with the closest WRF grid point.
Because WRF has much higher horizontal resolution, this collocation significantly undersamples WRF data, retaining only ca.~1\% of all WRF grid points.
However, because WRF's effective resolution is $\sim7\Delta x$ \citep{skamarock2021}, i.e., $\sim14$\,km for this study, the undersampling yields a more independent dataset with less redundant information.
Vertically, we retain only the lower 30 pressure levels from ERA5, as our WRF simulations stop at 10\,hPa (ca.~20\,km height), whereas ERA5 extends up to 1\,hPa (ca.~30\,km height).

\paragraph*{Features}
The feature and target variables utilized in this study are summarized in tab.~\ref{tab:var_overview}.
We consider $p=9$ vertical profiles and $q=16$ single-level and surface variables, which are known to modulate the primary regression target $C_n^2$.
Since (optical) atmospheric turbulence is modulated by wind shear and buoyancy \citep{stull1988}, the vertical profiles of wind components $u$ and $v$ as well as potential temperature $\theta$ are included.
As it is not the absolute values of $u$, $v$, and $\theta$ but their vertical gradients that drive turbulence, we additionally compute the mean wind shear $S$ and the potential temperature gradient $\Gamma$ between adjacent levels as 
\begin{equation}
   S = {\left({(\Delta u / \Delta z)}^2 + {(\Delta v / \Delta z)}^2\right)}^{1/2}
   \label{eq:shear}
\end{equation}
and
\begin{equation}
   \Gamma = \Delta\theta/\Delta z.
   \label{eq:gamma}
\end{equation}
We mark these engineered features with (*) in tab.~\ref{tab:features}.

We introduce height information for each vertical level by including the pressure $p$ at each level and the corresponding height above ground in linear space ($z$) and in scaled log-space ($\textrm{lp}(z)$, cf.~eq.~\eqref{eq:lz}).
Additionally, the valid-pressure mask classifying levels as below or above ground (cf.~sec.~\ref{sec:sqf}) is included.

As for the profiles, we select surface variables related to wind shear and buoyancy.
These variables include the friction velocity $u_*$, the sensible and latent heat fluxes $Q_H$ and $Q_E$, and the W71-based estimate of the surface-layer $C_T^2$ (cf.~eq.~\eqref{eq:cn2_w71}).
To let the model learn diurnal and seasonal patterns, we include temporal features based on the hour of the day (hr) and the day of the year (doy).
Because they are periodic, we encode hr and doy using sine and cosine components.
Both features are also normalized to a $2\pi$ period as $\text{hr}' = (2\pi\,\text{hr})/24$ and $\text{doy}' = (2\pi\,\text{doy})/365$ before computing the sine and cosine.

\paragraph*{Target variables}
The $r=4$ target variables listed in tab.~\ref{tab:targets} are the $\log_{10}$-transformed $C_n^2$ from WRF at the 100 vertical levels and the physical variables used to parameterize $C_n^2$ (cf.~eq.~\eqref{eq:ct2_corrsin}): turbulent kinetic energy (TKE) $e$, potential temperature variance $\sigma_\theta^2$, and master length scale $L_m$.
We incorporate these additional targets as auxiliary variables to support model training, but this study focuses on predicting $C_n^2$.

\paragraph*{Normalization}
All features and targets are normalized to the range $[0, 1]$ before training.
The method for determining the normalization bounds differs between features and targets as indicated in tab.~\ref{tab:var_overview}.
For well-behaved variables with clear minimum and maximum values, such as height above ground or pressure, the absolute minimum and maximum values across all vertical columns (time and location) are used.
For variables with occasional outliers, such as $u$ and $v$, we use the 1st and 99th percentiles instead to avoid placing too much emphasis on outliers.
Note that normalization does not clip values, so values outside the normalization bounds still contribute to training.
If variables such as $u_*$ or $S$ have physically meaningful lower bounds at zero, these are enforced during normalization.

\paragraph*{Dataset summary}
Following the decoupled training approach described in sec.~\ref{sec:decoupled_training}, three datasets are utilized in this work.
First, the HR WRF dataset with 100 vertical levels serves as the training target.
Second, we use the WRF-PL dataset, a coarsened version of the WRF data with 30 pressure levels, as the LR input during training to emulate ERA5.
Third, the ERA5-PL dataset, also with 30 pressure levels, is used as the LR input for inference.

The following section presents results from experiments that systematically evaluate the Squeezeformer across these data configurations, isolating the contributions of regression, super-resolution, and distributional shift to overall prediction performance.

\begin{table*}
   \caption{Overview of experiments differing by performed tasks, data utilized for training and inference, and if output quantile mapping (oQM) was applied.}%
   \label{tab:experiments}
   \centering
   \begin{tabular}{lcc r@{\:}>{$}c<{$}@{\:}l cc}
      \toprule
      Experiment & \makecell{Regression\\ $f(\mathbf{X}, \mathbf{x}_s; \mathbf{z}_y) \approx \mathbf{Y}$} & \makecell{Superresolution \\ $m \to n$} & \multicolumn{3}{c}{\makecell{Training data\\ $(\mathbf{X}, \mathbf{x}_s) \to \mathbf{Y}$}} & \makecell{Inference data\\ $f(\mathbf{X}, \mathbf{x}_s; \mathbf{z}_y)=\hat{\mathbf{Y}}$} & oQM \\
      \cmidrule(lr){1-1} \cmidrule(lr){2-3} \cmidrule(lr){4-7} \cmidrule(lr){8-8}
      Ctl. & \checkmark & -- & WRF HR & \to & WRF HR & WRF HR & -- \\
      WRF-PL & \checkmark & 30 $\to$ 100 & WRF-PL & \to & WRF HR & WRF-PL & -- \\
      ERA5-PL oQM & \checkmark & 30 $\to$ 100 & \multicolumn{3}{c}{same as \emph{WRF-PL}} & ERA5-PL & \checkmark \\
      \midrule
      HV+W71 & \checkmark & 30 $\to$ 100 & \multicolumn{3}{c}{\makecell{$W$ from\\WRF-PL}} & \makecell{$z$ from\\ WRF HR} & -- \\
      \bottomrule
   \end{tabular}
\end{table*}

\section{Results}\label{sec:results}
This section presents the results of experiments conducted to assess the performance of the Squeezeformer model in estimating high-resolution $C_n^2$ profiles from low-resolution meteorological data.
The experiments are summarized in tab.~\ref{tab:experiments} and are set up to increase in complexity step by step to illuminate different aspects of the model's performance.
The simplest experiment, the control run \emph{Ctl.}, uses the native HR WRF data as inputs and as targets.
Taking only the features listed in tab.~\ref{tab:features}, the \emph{Ctl.}\ model performs regression toward $C_n^2$ but does not perform super-resolution or account for spatio-temporal shifts.
Its purpose is to show that $C_n^2$ can be estimated from the given inputs and to serve as an upper-performance baseline for the subsequent analysis.
The second experiment, \emph{WRF-PL}, uses WRF-PL data as input and HR WRF data as the target, performing both regression and super-resolution.
This experiment is key to our decoupled training approach, in which the WRF-PL dataset aims to emulate ERA5-PL without spatiotemporal misalignment.
The trained \emph{WRF-PL} model is then used in the third experiment, \emph{ERA5-PL oQM}, where predictions are made from ERA5-PL data with quantile-mapping-corrected outputs to simulate the operational situation in which WRF-PL data are not available.
The final experiment in tab.~\ref{tab:experiments} is the lower baseline, in which the HV+W71 model is evaluated on WRF-PL data.
In Appendix~A, we demonstrate that statistically aligning the ERA5-based outputs to WRF HR data using oQM improves prediction accuracy.
That appendix also compares our proposed decoupled training approach with directly training a Squeezeformer on ERA5-PL inputs and WRF HR targets, showing that the decoupled approach yields more realistic predictions that better reflect the vertical structure of the $C_n^2$ profiles.
Additionally, we apply the trained model without retraining to a new domain over the Canary Islands to assess how well OTProf transfers to a different climate and complex terrain (cf.~tab.~\ref{tab:cn_experiments}).

The Squeezeformer models in all experiments are configured and evaluated identically.
The $p+q=25$ input features (cf.~tab.~\ref{tab:features}) are expanded to $d=48$ embedding dimensions in the first layer before being processed by 4 Squeezeformer blocks.
To facilitate the extraction of small-scale features from the input profiles, a convolutional kernel size of 3 is used in the convolutional layers of the Squeezeformer blocks.
Each transformer block has 2 attention heads, and the regression space has size $h=512$.
The prediction and confidence heads map this high-dimensional space back to the $r=4$ target variables.
For robust training, the Squeezeformer is trained on $n_q=32$ randomly sampled target levels, but all $n=100$ levels are used for testing and the results presented in the following.
The resulting model has 1 million trainable parameters. 
The temporal staggering of the WRF simulations allows us to split the DL dataset into training, validation, and test sets by selecting different years.
For training, data from 2019 and 2020 are used, while 2018 is reserved for validation and 2017 for testing.
All datasets cover all seasons to ensure statistical representativeness, but being from different years avoids data leakage.

The \emph{WRF-PL} Squeezeformer model was trained on three NVIDIA RTX A6000 GPUs for 600~epochs, taking 2\,h, resulting in a total training cost of 6 GPU hours.
Inference on the test data (ca.\ 90~days at ca.\ 200~ERA5 grid points in the NL domain) takes 12\,s on a single RTX A6000 GPU.
Consequently, a full year of high-resolution profiles over the Netherlands can be produced in under 1\,min of GPU time or a 30-year $C_n^2$ climatology in less than 30\,min.

The results presented below are based on the independent 2017 test set.
We begin with a visual comparison of predicted and reference $\log_{10} C_n^2$ profiles, followed by a statistical quantification of agreement in sec.~\ref{sec:res_lcn2_stats}.
Using a structure function (SF) analysis, we examine how well different models capture the vertical structure of $C_n^2$.
We discuss the practical impact of the performance differences between models in the context of the integrated astroclimate parameters $r_0$ and $\sigma_I^2$ in sec.~\ref{sec:res_r0_si2}.
Finally, sec.~\ref{sec:res_transfer} presents the out-of-domain study for the Canary Islands.

\subsection{Estimated $\log_{10}C_n^2$ profiles}\label{sec:res_lcn2_stats}
\begin{figure*}[t]
   \centering
   \includegraphics[width=.9\textwidth]{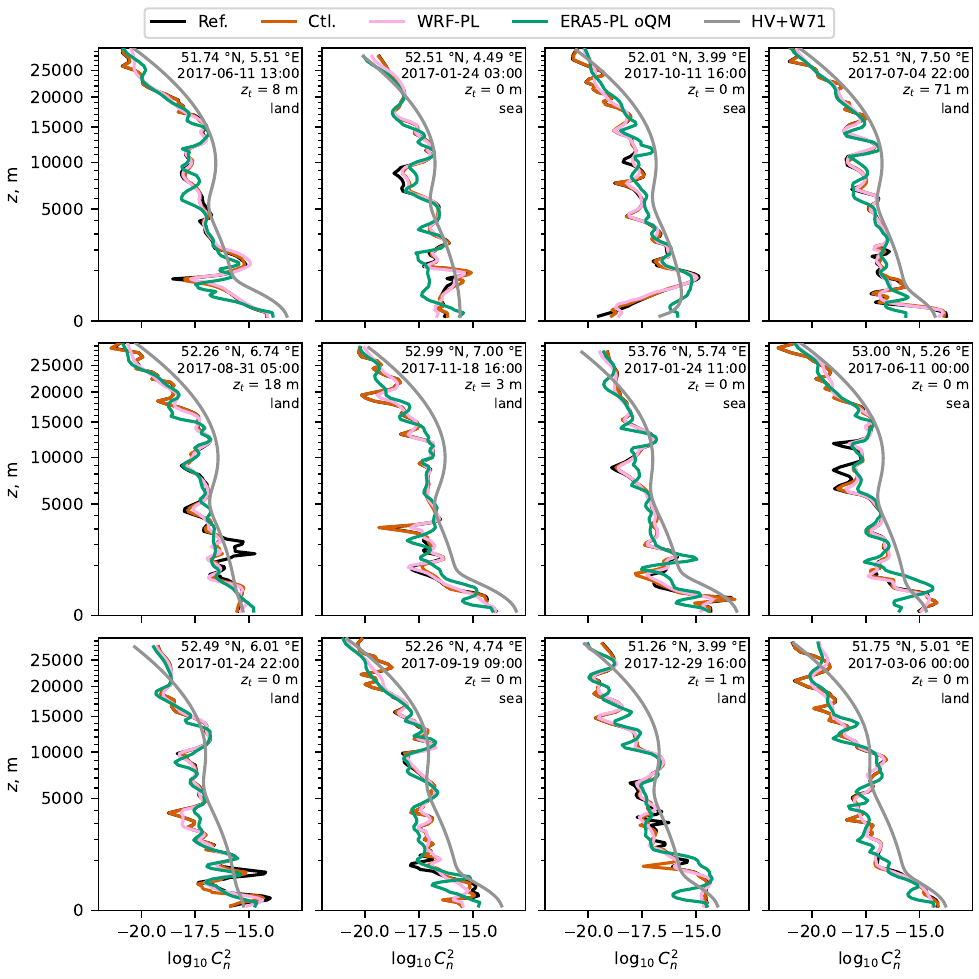}
   \caption{Randomly drawn examples of $\log_{10} C_n^2$ profiles from test set for qualitative visual performance assessment. Location, time, terrain height $z_t$, and land/sea classification for each profile are given in the panels.}\label{fig:lcn2_profiles}
\end{figure*}

\begin{figure*}
   \centering
   \begin{subfigure}[t]{.48\textwidth}
      \caption{Scores for estimated $\log_{10}C_n^2$. Arrows indicate direction of improvement for each score. The numerical values are given in tab.~\ref{tab:app_main_lcn2}.}%
      \label{fig:lcn2_scores}
      \includegraphics[width=\textwidth]{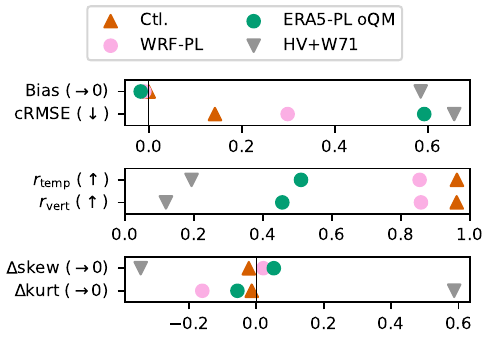}%
   \end{subfigure}%
   \hspace{2em}%
   \begin{subfigure}[t]{.40\textwidth}
      \caption{Temporal and vertical correlation between WRF-PL and ERA5-PL model inputs.}%
      \label{fig:input_agreement}
      \includegraphics[width=.75\textwidth]{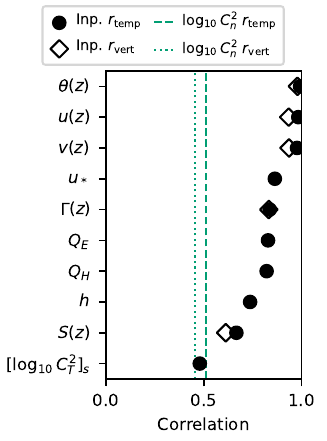}%
   \end{subfigure}
   \caption{Performance scores for $\log_{10}C_n^2$ estimates for the four experiments in panel (a), with correlation between WRF-PL and ERA5-PL input for reference in panel (b).}%
   \label{fig:lcn2_scores_input}
\end{figure*}

A visual impression of the performance of the different models is given in fig.~\ref{fig:lcn2_profiles} with scores for the full test set shown in fig.~\ref{fig:lcn2_scores}.
The panels in fig.~\ref{fig:lcn2_profiles} display $\log_{10} C_n^2$ profiles at randomly selected time instances and locations, where the reference HR WRF profiles are shown in black and the predictions of the four experiments with \emph{Ctl.}\ in red, \emph{WRF-PL} in pink, \emph{ERA5-PL oQM} in green, and \emph{HV+W71} in grey.
The upper baseline's \emph{Ctl.}\ profiles closely resemble the reference profiles and are often indistinguishable from the reference profiles.
The model captures large-scale trends across the atmospheric boundary layer (ABL) and local small-scale features confined to a few 100\,m, with only a few underestimations of sharp edges relative to the ground truth.
The performance scores reflect this visual observation with a bias close to zero, small cRMSE, and very high vertical and temporal correlation scores. 
The \emph{HV+W71} profiles, representing the lower baseline, surprisingly match the general trend.
However, they clearly miss the small-scale features present in all other profiles, as reflected by the low $r_\textrm{vert}$, and consistently overestimate the reference ($\log_{10}C_n^2$ bias of 0.584).
In conclusion, the \emph{HV+W71} and \emph{Ctl.}\ predictions serve as effective and simple upper and lower baselines before super-resolution and spatio-temporal misalignment are introduced as additional challenges in the \emph{WRF-PL} and \emph{ERA5-PL oQM} experiments.

The profiles (cf.~fig.~\ref{fig:lcn2_profiles}) and the scores (cf.~fig.~\ref{fig:lcn2_scores}) show that the \emph{WRF-PL} predictions are visually and statistically situated between the upper baseline \emph{Ctl.}\ and the lower baseline \emph{HV+W71}.
\emph{WRF-PL} exhibits greater smoothing than \emph{Ctl.}, but captures the reference profiles well, given the significant super-resolution task from 30 to 100 levels.
Similar to \emph{Ctl.}, \emph{WRF-PL} shows almost no bias but has reduced vertical and temporal correlation as well as higher cRMSE due to smoothing and occasionally misplaced features.
Nevertheless, temporal and vertical correlation values above 0.8 are still very satisfactory and demonstrate that the Squeezeformer architecture is suited for the task.

The \emph{ERA5-PL oQM} profiles and statistics show larger disagreement with the reference compared to the WRF-based estimates.
While the bias of the estimates is close to zero due to the oQM, the cRMSE doubles compared to the \emph{WRF-PL} experiments. 
Also, temporal and vertical correlation with the WRF reference drop significantly to $r_\textrm{temp}=0.51$ and $r_\textrm{vert}=0.46$.
However, we argue that this apparent performance decrease largely stems from the inherent misalignment between ERA5 and WRF data.
To put the performance scores into perspective, panel (b) of fig.~\ref{fig:lcn2_scores_input} presents temporal and vertical correlation between the ERA5-PL and WRF-PL \emph{model inputs}.
The ERA5-PL \emph{output} correlation values from panel (a) are shown as vertical lines for reference.
While wind and temperature correlate very well in time and height, the inputs expected to modulate turbulence correlate less well.
The surface fluxes $u_*$, $Q_H$, and $Q_E$ show reduced temporal correlation of $\sim0.8$, just like the potential temperature gradient $\Gamma$.
Wind shear $S$ and W71-based surface $\log_{10}C_T^2$ are impacted more strongly with $r$ values around 0.7 and 0.5, respectively.
As our decoupled approach explicitly does not aim to compensate for such timing and vertical placement differences, it is expected that low correlations in the inputs carry over into the predictions.
Instead, we focus on representing vertical structure and the distribution of values well, which is discussed in more detail in the following.

Nevertheless, compared to \emph{HV+W71}, \emph{ERA5-PL oQM} still shows clear visual and quantitative improvements despite the spatio-temporal misalignment.

\begin{figure*}
   \centering
   \begin{subfigure}[t]{\textwidth}
      \caption{Probability density functions and quantile-quantile plots of $\log_{10} C_n^2$ predictions against WRF-based reference profiles.}%
      \label{fig:lcn2_stats_dist}
      \centering
      \includegraphics[width=.8\textwidth]{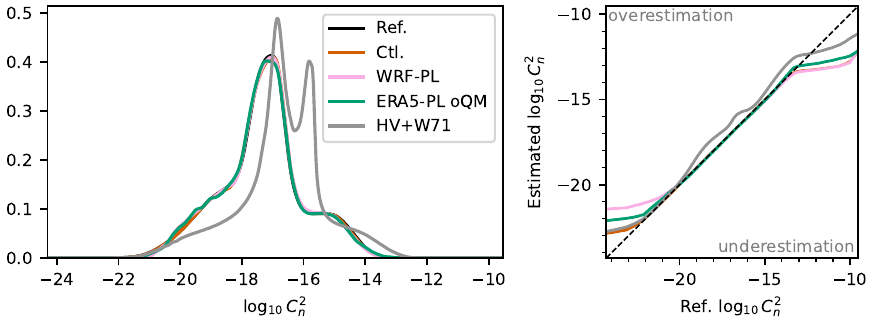}
   \end{subfigure}
   \begin{subfigure}[t]{\textwidth}
      \caption{Vertical second-order structure functions of $\log_{10} C_n^2$ profiles.}%
      \label{fig:lcn2_sf}
      \includegraphics[width=.9\textwidth]{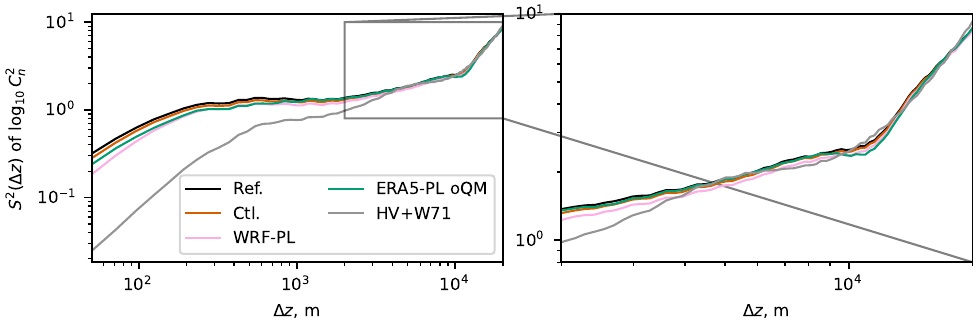}
   \end{subfigure}
   \caption{Statistical characteristics of predicted $\log_{10} C_n^2$ profiles compared to WRF-based reference profiles. All curves are based on data aggregated over all levels, locations, and time.}%
   \label{fig:lcn2_stats_sf}
\end{figure*}

\paragraph*{Statistical characteristics}
Figure~\ref{fig:lcn2_stats_sf} summarizes the statistical characteristics of the model predictions relative to the WRF reference profiles.
Panel (a) compares the distributions of $\log_{10} C_n^2$ in two ways, with probability density functions (PDFs) on the left and quantile-quantile (qq) plots on the right.
The qq plots compare the empirical density functions of the predictions and the reference profiles, where a perfect match would yield a straight 1:1 line.
The shape differences of the estimated PDF compared to the reference are quantified as skewness differences $\Delta \textrm{skew}$ and excess kurtosis difference $\Delta \textrm{kurt}$ in fig.~\ref{fig:lcn2_scores}.
Panel (b) shows the vertical second-order SFs of $\log_{10} C_n^2$ profiles.
Contrary to the previous discussion, the global statistical characteristics are not affected by spatio-temporal misalignment, as no profile-by-profile comparison is made.
Instead, the statistical characteristics reflect how well different models produce profiles with the correct vertical structure and magnitude.

The PDF of \emph{Ctl.}\ matches the reference very well, consistent with the high scores reported earlier. 
However, deviations become visible in the qq-plot in the tails of the distribution. 
Very low values ($10^{-22}$\,m$^{-2/3}$) are overestimated, likely because regularization during training prevents overfitting to the few potential outliers in this range.
A pronounced kink is visible at high values ($C_n^2 \approx 10^{-13}$\,m$^{-2/3}$), indicating underestimation of the high tail.
The underestimation of the high tail is more critical than the low-tail overestimation, as high $C_n^2$ values are most relevant for FSOC applications because they can cause link degradation or outages.
As observations are typically less smooth than WRF-based $C_n^2$ estimates, this underestimation is likely more pronounced against observations, which should be assessed in future work.
On the other hand, these very strong conditions make up less than 0.01\% of the full WRF reference data (0.11\% below 100\,m), so while capturing them well is desirable, we consider the performance of \emph{Ctl.}\ good for the purpose of this study.

\emph{WRF-PL} and \emph{ERA5-PL oQM} track the reference PDFs similarly well as \emph{Ctl.}
Because only the ERA5-based predictions are bias-corrected, \emph{ERA5-PL oQM} captures the distribution tails better than \emph{WRF-PL}, as shown by later deviations of the qq-plot from the 1:1 line.
A lower $\Delta \textrm{kurt}$ for \emph{ERA5-PL oQM} reflects this, while the skewness is slightly more overestimated.
Compared to \emph{Ctl.}, both distributions overestimate the lower tail more, which is acceptable because very weak turbulence conditions have low practical relevance.
Overall, also the ERA5-based $ C_n^2$ estimates capture the bulk of the distribution very well.
The similarity between the two PL experiments supports the conclusion that quantile-mapping ERA5-PL outputs to WRF $C_n^2$ is effective.
In Appendix~A, we also show that uncorrected ERA5-PL outputs lead to stronger overestimation of the lower tail.
These results also support the decoupled training approach, demonstrating that statistically consistent estimates can be obtained even when using ERA5-PL data for inference.
The \emph{HV+W71} model, by contrast, does not need much detailed consideration. 
Its simple analytical formulation does not capture the statistical characteristics well, as shown by the shifted bimodal distribution and mismatched qq-plot.

\paragraph*{Representation of vertical structure}
The capability of the different experiments to capture the vertical structure of $C_n^2$ is assessed by the SF in fig.~\ref{fig:lcn2_sf}.
Assessing how well the vertical structure is captured matters because smoothing small-scale features can lead to overestimating (or underestimating) $r_0$ ($\sigma_I^2$), as discussed in sec.~\ref{sec:res_r0_si2}.
The \emph{Ctl.}\ curve follows the WRF-based reference SF closely over the entire range of scales with only a small widening gap toward smaller scales.
This gap reflects the occasional smoothing of local/small-scale features observed in fig.~\ref{fig:lcn2_profiles}, but overall vertical structure is reproduced very well.
The \emph{HV+W71} model, on the other hand, shows clear deficits in capturing vertical structure, as was already evident from visual inspection and $r_\textrm{vert}$.
The \emph{HV+W71} SF matches the reference at very large scales, consistent with the visual impression that \emph{HV+W71} reflects the overall trend of $C_n^2$.
However, the \emph{HV+W71} SF quickly and strongly departs from the WRF-based reference toward smaller scales, indicating oversmoothing.
As expected, this observation follows from the simple analytical nature of the \emph{HV+W71} model.

As before, \emph{WRF-PL} and \emph{ERA5-PL oQM} track the reference SF closely, although they show widening gaps toward smaller scales.
These gaps correspond to the small-scale smoothing observed visually earlier.
At large scales ($\Delta z>1$\,km), the curves are almost indistinguishable from the reference, and only at smaller scales do the PL-based models start deviating.
Interestingly, \emph{WRF-PL} loses variance at very small scales faster than the \emph{ERA5-PL oQM} predictions, but the deviation is too small to yield useful insights.
In Appendix~A, we show that ERA5-based predictions without oQM lack overall variance, resulting in a structure function with a constant offset, which oQM fixes.
In summary, all Squeezeformer-based experiments remain substantially closer to the WRF reference than \emph{HV+W71}, indicating good overall performance and well-captured vertical structure.

\begin{figure*}[t]
   \begin{subfigure}[t]{\textwidth}
      \centering
      \caption{Fried parameter $r_0$}%
      \label{fig:r0}
      \begin{minipage}[t]{.35\textwidth}
         \vspace{-\topskip}
         \includegraphics[width=\textwidth]{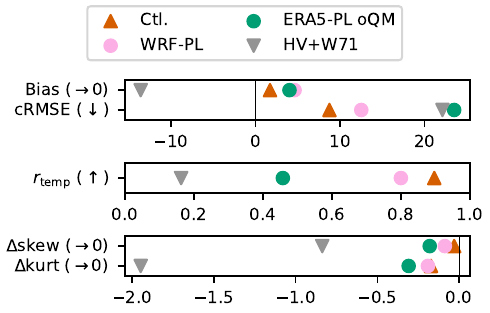}%
      \end{minipage}
      \hfill
      \begin{minipage}[t]{.64\textwidth}
         \vspace{-\topskip}
         \vspace{.5em}
         \includegraphics[width=\textwidth]{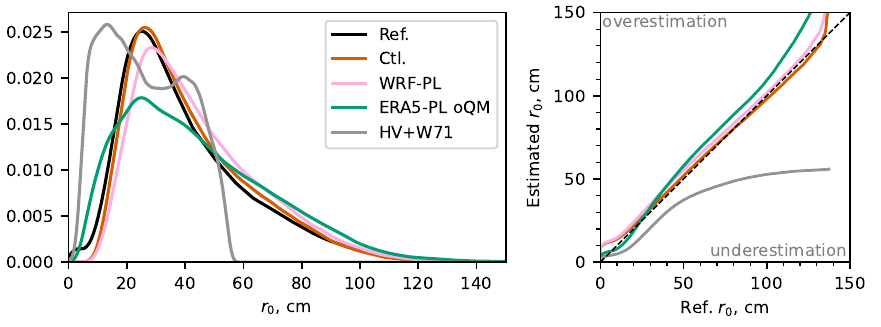}
      \end{minipage}
   \end{subfigure}

   \begin{subfigure}[t]{\textwidth}
      \centering
      \caption{Logarithm of scintillation index $\log_{10}\sigma_I^2$}%
      \label{fig:si}
      \begin{minipage}[t]{.35\textwidth}
         \vspace{-\topskip}
         \includegraphics[width=\textwidth]{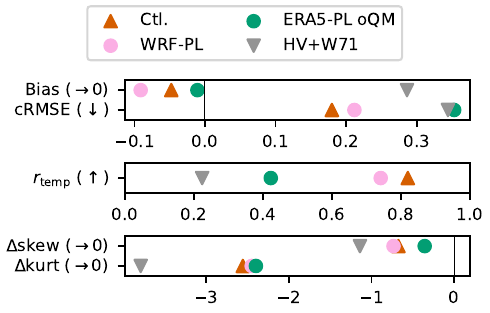}%
      \end{minipage}
      \hfill
      \begin{minipage}[t]{.64\textwidth}
         \vspace{-\topskip}
         \vspace{.7em}
         \includegraphics[width=\textwidth]{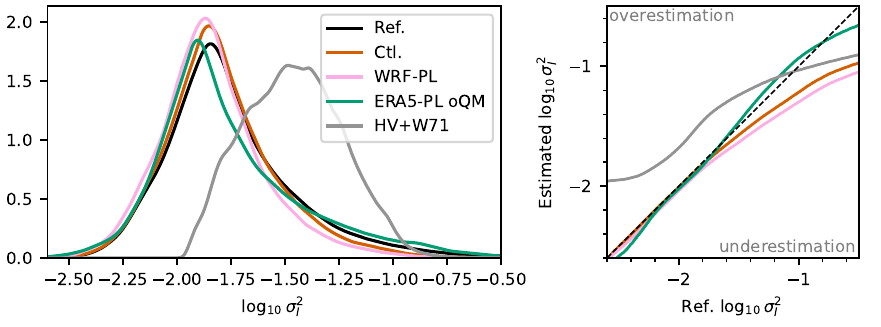}
      \end{minipage}
   \end{subfigure}
   \caption{Performance scores, histograms, and quantile-quantile plots for the integrated OT parameters $r_0$ and $\sigma_I^2$ computed from the predicted $C_n^2$ profiles. Numerical scores are given in tabs.~\ref{tab:app_main_r0} and \ref{tab:app_main_lsi2}.}%
   \label{fig:r0_si2}
\end{figure*}

Overall, the statistical assessment and the SF study demonstrate that OTProf can yield accurate and physically-consistent estimates of $C_n^2$ profiles, even with ERA5-PL inputs.
The input correlation analysis (cf.~fig.~\ref{fig:input_agreement}) suggests that the drop in correlation performance is largely due to mismatches in the input data, but future work should further disentangle the error source.
Therefore, OTProf with ERA5-PL data enables the temporal extrapolation of a limited set of WRF-based $C_n^2$ to generate regional climatologies.
Next, we assess how column-integrated parameters based on the estimated profiles compare with their references.

\subsection{Estimates of $r_0$ and $\sigma_I^2$}\label{sec:res_r0_si2}

After assessing the predicted $C_n^2$ profiles, we discuss how observed smoothing and distributional deviations affect the integrated astroclimate parameters $r_0$ and $\sigma_I^2$. 
In analogy to the profile analysis, performance scores and distributions for $r_0$ and $\sigma_I^2$ computed from the predicted $C_n^2$ profiles are presented in figs~\ref{fig:r0} and \ref{fig:si}, respectively.
Before proceeding, we note that the Fried parameter $r_0$ and $C_n^2$ are inversely related: low $r_0$ corresponds to high turbulence/$C_n^2$, and vice versa, whereas the scintillation index $\sigma_I^2$ increases with turbulence and $C_n^2$.

\paragraph*{Fried Parameter}
Considering $r_0$ in fig.~\ref{fig:r0}, the \emph{Ctl.}\ experiment again performs best with high temporal correlation and low bias, cRMSE and shape differences compared to the reference.
The distributions also match well across most of the range, with only small overestimations for small ($r_0 < 20$\,cm) and large ($r_0 > 130$\,cm) values.
Interestingly, no underestimation of $r_0$ is observed in the qq-plot, which can be explained by linking the $r_0$ analysis back to the earlier $C_n^2$ profile assessment.
When $C_n^2$ is overestimated, turbulence is too high, so $r_0$ should be too low, i.e., underestimated.
However, the $C_n^2$ overestimations discussed previously occurred for very small values in log-space, whereas $r_0$ is based on the integral of $C_n^2$ in linear space.
Consequently, overestimating low $C_n^2$ values has a negligible effect on the integrated $r_0$, which explains why no underestimation appears in the qq-plot.
The opposite case reveals the impact more clearly: when high $C_n^2$ values are underestimated, the error in linear space becomes very pronounced, leading to overestimated $r_0$.
This overestimation is attributed to the smoothing of previously observed profiles, both in the structure function analysis and visually in fig.~\ref{fig:lcn2_profiles}.

The behavior of \emph{WRF-PL} and \emph{ERA5-PL oQM} is similar to \emph{Ctl.}\ in that they also show skewed single-mode distributions with long tails toward high $r_0$.
However, the distributions of \emph{WRF-PL} and \emph{ERA5-PL oQM} are progressively less skewed and with progressively more underestimated high tails.
This is visible in the qq plot and in the progressively decreasing $\Delta \textrm{skew}$ and $\Delta \textrm{kurt}$ values.
While \emph{WRF-PL} is still quite close to and similar to the reference, the \emph{ERA5-PL oQM} $r_0$ distribution is notably wider, which is also reflected in its larger cRMSE.
The widening results in turbulence conditions with $r_0<15$\,cm and $r_0>45$\,cm appearing too often, while the probability density in between is too low.
We attribute this behavior to a remaining mismatch between ERA5 and WRF.
While Appendix~A shows that oQM on $\log_{10}C_n^2$ also improves the bias and shape of the $r_0$ distribution, it is a global adjustment applied to the full column.
In reality, stronger bias corrections may be needed near the surface, where surface effects likely introduce the largest mismatch between datasets, compared to the free atmosphere, which is dominated by well-resolved large-scale phenomena.
While future research into more fine-grained dataset alignment techniques is recommended, we are still satisfied by the generally good performance of the ERA5-based $r_0$ estimates -- especially when compared against \emph{HV+W71}.

\paragraph*{Scintillation Index}
The performance characteristics of $\sigma_I^2$ in fig.~\ref{fig:si} differ from those of $r_0$ because the integrand uses height-dependent weighting, whereas $r_0$ uses uniform weighting.
Since turbulence strength and $C_n^2$ generally decrease with height, underestimating high $C_n^2$ near the surface has less impact on $\sigma_I^2$.
All DL-based distributions are visually close to the reference, while \emph{HV+W71} exhibits a large bias and deviation in shape.
The qq plots of the \emph{Ctl.}\ and \emph{WRF-PL} experiments show good agreement up to $\sigma_I^2 \approx 10^{-1.8}$ where they begin to underestimate the strong turbulence tail.
It is surprising that also \emph{Ctl.}\ shows this behavior.
The reason could be insufficient representation in the training data or architecture limitations, which are left for future work to investigate.
The oQM bias correction helps \emph{ERA5-PL oQM} to keep $\log_{10}\sigma_I^2$ closer to the 1:1 line for longer than \emph{Ctl.}\ and \emph{WRF-PL} but overcorrects for strong turbulence, leading to overestimated scintillation.
Due to the height-dependent weighting, we assume that too large a high-altitude $C_n^2$ is responsible for the overshoot, again suggesting that the bulk oQM should be refined in future work.

\paragraph*{FSOC Implications}
The $r_0$ and $\sigma_I^2$ analysis shows that the ERA5-based estimates reproduce the distributions of both parameters much better than \emph{HV+W71}.
Their temporal correlation with the WRF reference is lower than for the WRF-based experiments, with $r_\textrm{temp}=0.46$ for $r_0$ and $r_\textrm{temp}=0.42$ for $\log_{10}\sigma_I^2$.
As discussed in sec.~\ref{sec:res_lcn2_stats}, this reduction could largely stem from mismatches between ERA5 and WRF inputs, but future comparisons with observations should assess whether this is an artifact of comparing mismatched datasets or a model limitation.
For now, we recommend using the ERA5-based estimates for statistical and climatological FSOC assessments, e.g., link availability or site selection.
The following discussion therefore focuses on the distributions of $r_0$ and $\sigma_I^2$.

For a space-to-ground communication terminal with telescope aperture diameter $D$, the turbulence-induced degradation of the optical wavefront is determined by the ratio $D/r_0$ \citep{andrews2005,Sasiela2007}.
More specifically, degradation occurs for $D/r_0 > 1$, or $r_0 < D$.
Space-to-ground terminals typically have aperture diameters between 50\,cm and 100\,cm (e.g., \citet{saathof2017,klop2021}), so degradations are expected for $r_0$ values below this range.
According to the reference distribution in fig.~\ref{fig:r0}, $r_0$ values below 100\,cm occur in approximately 98.2\% of cases, and values below 50\,cm occur in approximately 70.7\% of cases, whereas the \emph{ERA5-PL oQM} predictions suggest 97.2\% and 64.6\% of cases, respectively.
The ERA5-based estimates thus slightly underestimate how often degraded conditions occur.
This follows from the positive $r_0$ bias of 4\,cm and the wider distribution, which places too many cases at $r_0>45$\,cm.
The positive bias is consistent with the underestimation of $C_n^2>10^{-13}$\,m$^{-2/3}$ by all DL models (cf.~fig.~\ref{fig:lcn2_stats_dist}), as $r_0$ is most sensitive to strong near-surface turbulence.
The effect decreases with increasing aperture diameter, as large apertures are affected by turbulence in almost all cases.
At the same time, strong turbulence with $r_0<15$\,cm, i.e., $D/r_0>3$ for a 50\,cm aperture, appears slightly too often.
Consequently, \emph{ERA5-PL oQM} is slightly optimistic about how often a link is degraded, but slightly pessimistic about how often the degradation is severe.

The scintillation index $\sigma_I^2$ modulates link losses due to intensity fluctuations.
Following \citet{giggenbach2008}, the corresponding link loss is proportional to $\left(\sigma_I^2\right)^{2/5}$.
An error in $\log_{10}\sigma_I^2$ therefore reduces to 2/5 of its size in the logarithm of the link loss.
\emph{ERA5-PL oQM} follows the 1:1 line in the qq plot of fig.~\ref{fig:si} up to $\sigma_I^2 \approx 10^{-1.5}$, which covers the bulk of conditions.
For stronger scintillation, \emph{ERA5-PL oQM} overestimates $\sigma_I^2$ and thus leads to conservative link loss estimates.
In contrast, \emph{Ctl.}\ and \emph{WRF-PL} underestimate $\sigma_I^2$ in this range, which would lead to optimistic link loss estimates.
The most extreme scintillation values are underestimated by all DL models.
As communication outages are most likely to occur under these conditions, future work is needed to improve performance in the high scintillation regime.

In conclusion, OTProf with ERA5-PL inputs provides $r_0$ and $\sigma_I^2$ statistics that are useful for FSOC planning and much more realistic than those of \emph{HV+W71}.
The trained OTProf model can now temporally extrapolate long ERA5-PL records to build a reliable climatology of $C_n^2$ and derived integrated and FSOC-relevant parameters.
Still, further work is needed to improve performance under strong turbulence and to assess the practical relevance of the low correlation scores relative to the WRF reference.
As with any DL model, it is also not clear how well a trained model performs in a new domain, especially when climatology or terrain are very different.
While thorough spatial generalization studies are needed in the future, the following section will briefly analyze how transferable OTProf is to a new domain.

\subsection{Transfer to a new domain}\label{sec:res_transfer}
\begin{table*}[t]
   \caption{Overview of Canary Islands (CI) out-of-domain experiments. All Squeezeformer-based experiments reuse the model trained on the Dutch domain (NL) and only differ by the WRF $C_n^2$ data used for oQM.}%
   \label{tab:cn_experiments}
   \centering
   \begin{tabular}{lcc r@{\:}>{$}c<{$}@{\:}l cc}
      \toprule
      Experiment & \makecell{Regression\\ $f(\mathbf{X}, \mathbf{x}_s; \mathbf{z}_y) \approx \mathbf{Y}$} & \makecell{Superresolution} & \multicolumn{3}{c}{\makecell{Training data\\ $(\mathbf{X}, \mathbf{x}_s) \to \mathbf{Y}$}} & \makecell{Inference data\\ $f(\mathbf{X}, \mathbf{x}_s; \mathbf{z}_y)=\hat{\mathbf{Y}}$} & oQM \\
      \cmidrule(lr){1-1} \cmidrule(lr){2-3} \cmidrule(lr){4-7} \cmidrule(lr){8-8}
      no oQM & \checkmark & \checkmark & WRF-PL (NL) & \to & WRF HR (NL) & ERA5-PL (CI) & -- \\
      NL-oQM & \checkmark & \checkmark & \multicolumn{3}{c}{same as above} & same as above & \makecell{WRF HR (NL),\\$\sim180$\,days} \\
      CI-oQM & \checkmark & \checkmark & \multicolumn{3}{c}{same as above} & same as above & \makecell{WRF HR (CI),\\$\sim20$\,days} \\
      \midrule
      HV+W71 & \checkmark & \checkmark & \multicolumn{3}{c}{\makecell{$W$ from\\WRF-PL (CI)}} & \makecell{$z$ from\\ WRF HR (CI)} & -- \\
      \bottomrule
   \end{tabular}%
\end{table*}
\begin{figure*}[t]
   \centering
   \includegraphics[width=0.9\textwidth]{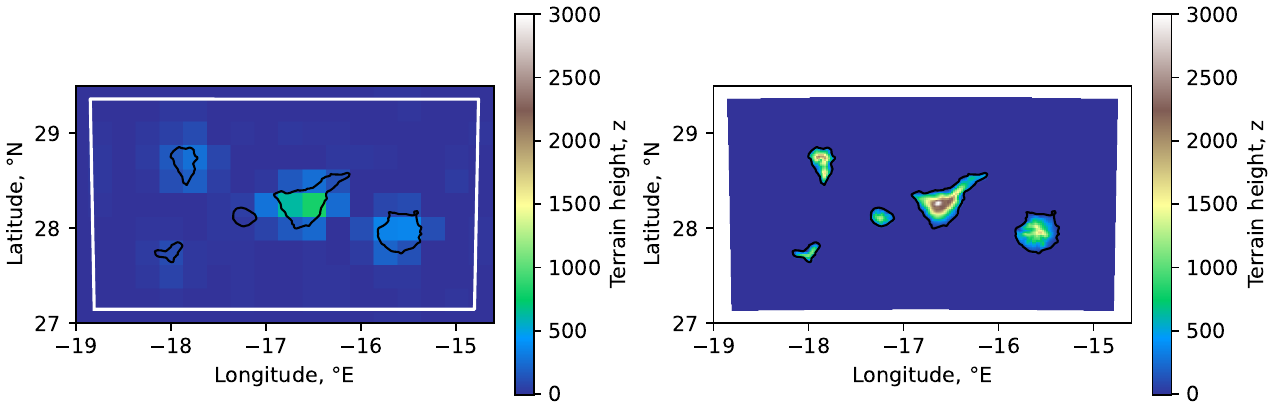}
   \caption{Elevation map of the Canary Islands from ERA5 (left) and WRF (right).}%
   \label{fig:domain_canaries}
\end{figure*}

The good performance of OTProf for the Dutch domain is promising, but producing the WRF simulations to train the model is still a substantial computational effort.
Therefore, this section briefly analyzes OTProf performance in a new domain and how to improve it when only a limited number of WRF simulations for that domain are available.
We would like to stress from the beginning that this out-of-domain analysis is only intended as an initial assessment and that more rigorous research in the future will be needed.

The region of interest for this out-of-domain (OOD) study covers the Canary Islands (CI) as visualized in fig.~\ref{fig:domain_canaries}.
The Canaries are selected for two reasons.
First, the climatology and terrain differ greatly from the Netherlands (NL), so the CI pose a challenge for the NL-trained model.
The climate is Mediterranean compared to the oceanic climate of the training domain~\citep{kottek2006}, and the volcanic islands have sloped terrain and elevations above 3000\,m above mean sea level, whereas the Dutch domain is mostly flat.
Second, La Palma and Tenerife each host astronomical observatories where many observational campaigns and modeling studies were carried out (e.g., \citet{munoz-tunon1998,masciadri2001,garcia-lorenzo2011,giordano2013,osborn2017}).
It is therefore a good benchmarking site for future detailed fine-tuning and performance studies with OTProf.

Following the setup described in sec.~\ref{sec:datasets}, we generate a small WRF-based $C_n^2$ dataset for validation.
We run twelve 5.5-day WRF simulations for 2017, the testing year for the Dutch domain.
The simulations start on the first day of each month, so all seasons are sampled.
Most of the domain is ocean, where OTProf is likely to perform better than in rough terrain.
To avoid overly optimistic results, we present results only for grid points classified as land.
As OTProf requires collocating the WRF columns with ERA5, it effectively operates at ERA5's coarse horizontal resolution, so the results are based on only eight land grid points.
We acknowledge that this reduces the statistical significance of our results, but we believe they remain insightful for assessing OOD behavior.

\begin{figure*}[t]
   \centering
   \begin{subfigure}[t]{\textwidth}
      \centering
      \caption{Randomly drawn examples of $\log_{10} C_n^2$ profiles.}%
      \label{fig:cn_land_lcn2_profiles}
      \includegraphics[width=.8\textwidth]{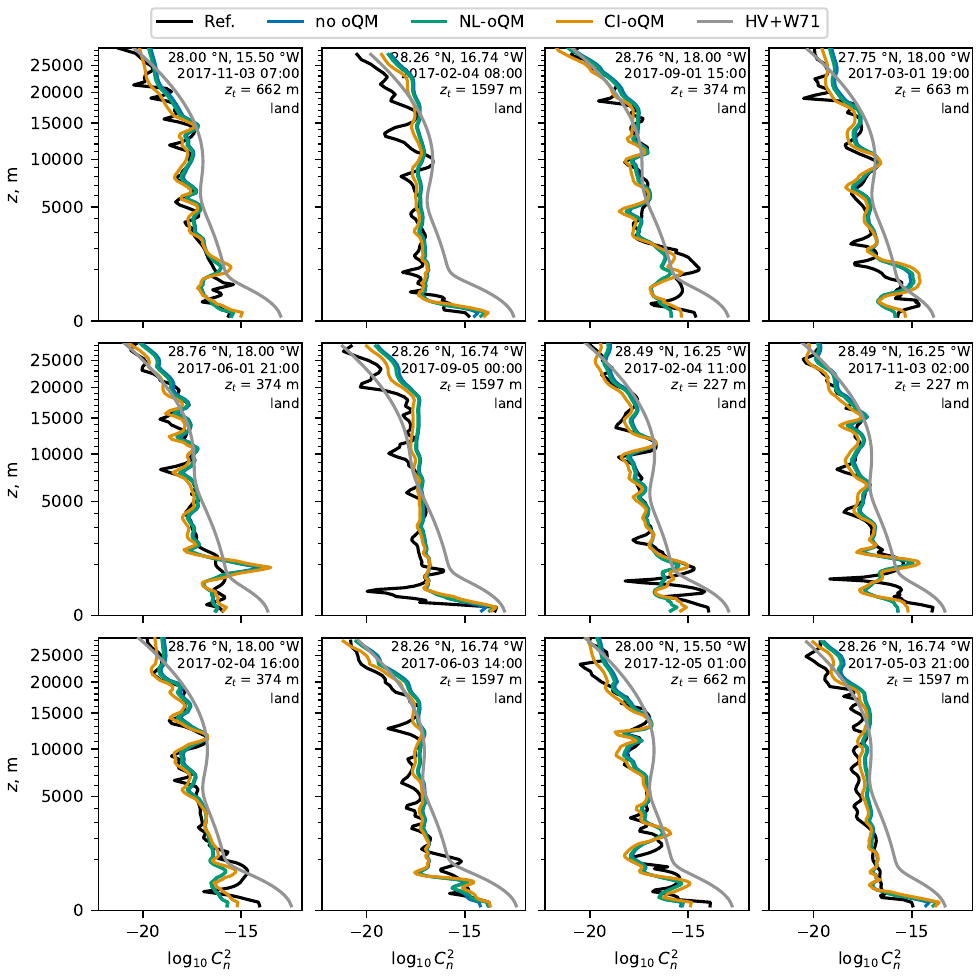}
   \end{subfigure}
   \begin{subfigure}[t]{\textwidth}
      \centering
      \caption{Vertical second-order structure functions of $\log_{10} C_n^2$ profiles, aggregated over all levels, land locations, and time.}%
      \label{fig:cn_land_lcn2_sf}
      \includegraphics[width=.8\textwidth]{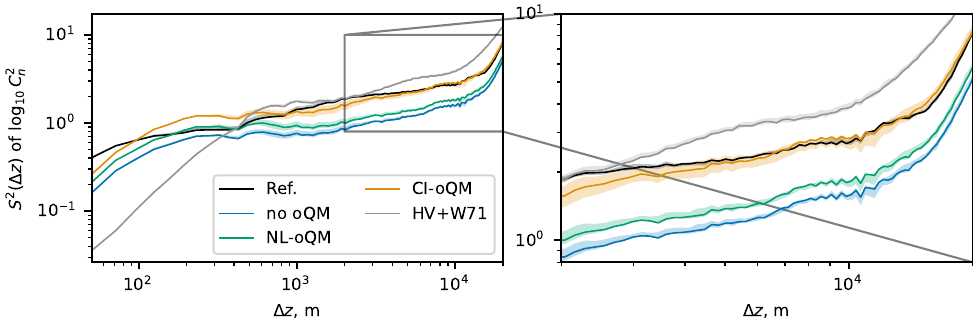}
   \end{subfigure}
   \caption{Canary Islands (out-of-domain test): examples and vertical structure of $\log_{10} C_n^2$ profiles over land. This figure is analogous to figs.~\ref{fig:lcn2_profiles} and \ref{fig:lcn2_sf}. ERA5-PL inputs are used for all predictions.}%
   \label{fig:cn_land_lcn2_profiles_sf}
\end{figure*}

\begin{figure*}[t]
   \centering
   \begin{subfigure}[t]{.48\textwidth}
      \caption{Scores for estimated $\log_{10}C_n^2$. Arrows indicate direction of improvement for each score.}%
      \label{fig:cn_land_lcn2_scores}
      \includegraphics[width=\textwidth]{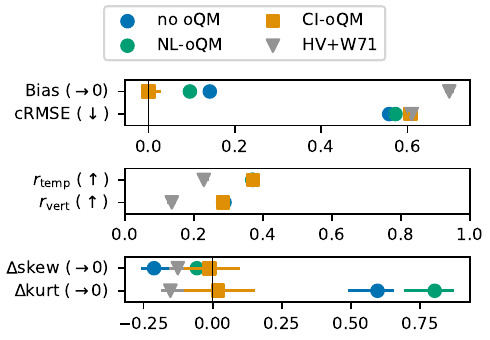}%
   \end{subfigure}%
   \hspace{2em}%
   \begin{subfigure}[t]{.40\textwidth}
      \caption{Temporal and vertical correlation between WRF-PL and ERA5-PL model inputs.}%
      \label{fig:cn_land_input_agreement}
      \includegraphics[width=.65\textwidth]{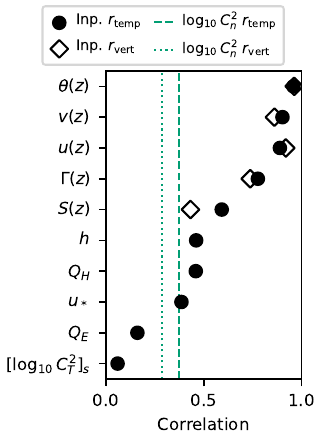}%
   \end{subfigure}

   \begin{subfigure}[t]{\textwidth}
      \centering
      \caption{Probability density functions and quantile-quantile plots of $\log_{10} C_n^2$ predictions against WRF-based reference profiles, aggregated over all levels, land locations, and time.}%
      \label{fig:cn_land_lcn2_stats_dist}
      \includegraphics[width=.8\textwidth]{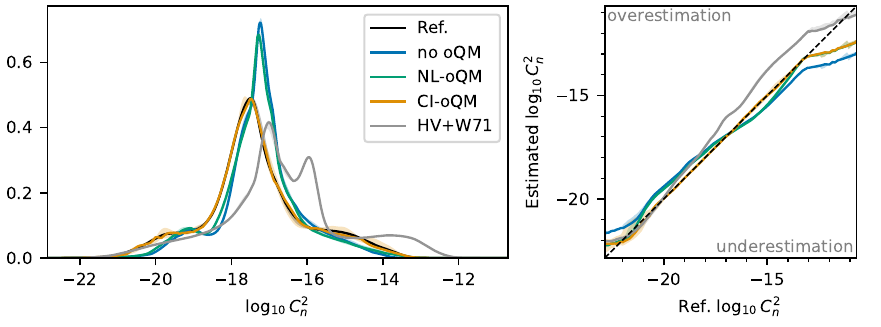}
   \end{subfigure}
   \caption{Canary Islands (out-of-domain test): performance scores and distributions of $\log_{10} C_n^2$ predictions over land. This figure is analogous to figs.~\ref{fig:lcn2_scores_input} and \ref{fig:lcn2_stats_dist}. Numerical scores are given in tab.~\ref{tab:app_cn_land_lcn2}. ERA5-PL inputs are used for all predictions.}%
   \label{fig:cn_land_lcn2_stats}
\end{figure*}

\begin{figure*}[t]
   \begin{subfigure}[t]{\textwidth}
      \centering
      \caption{Fried parameter $r_0$}%
      \label{fig:cn_land_r0}
      \begin{minipage}[t]{.35\textwidth}
         \vspace{-\topskip}
         \includegraphics[width=\textwidth]{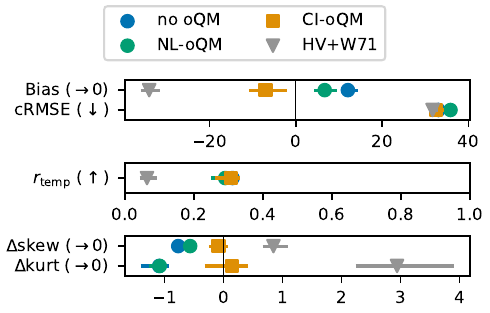}%
      \end{minipage}
      \hfill
      \begin{minipage}[t]{.64\textwidth}
         \vspace{-\topskip}
         \vspace{.4em}
         \includegraphics[width=\textwidth]{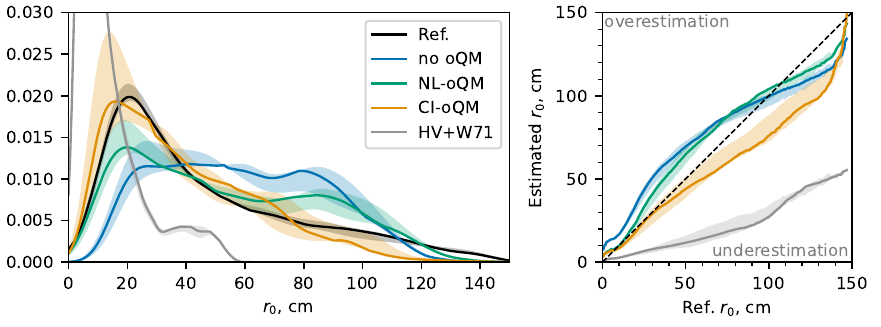}
      \end{minipage}
   \end{subfigure}

   \begin{subfigure}[t]{\textwidth}
      \centering
      \caption{Logarithm of scintillation index $\log_{10}\sigma_I^2$}%
      \label{fig:cn_land_si}
      \begin{minipage}[t]{.35\textwidth}
         \vspace{-\topskip}
         \includegraphics[width=\textwidth]{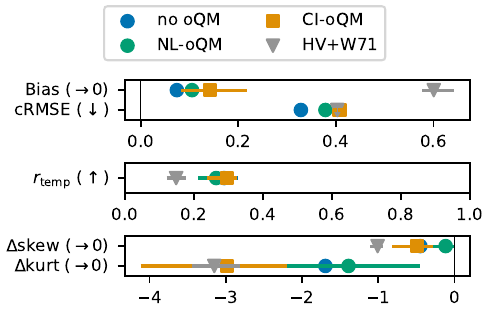}%
      \end{minipage}
      \hfill
      \begin{minipage}[t]{.64\textwidth}
         \vspace{-\topskip}
         \vspace{.4em}
         \includegraphics[width=\textwidth]{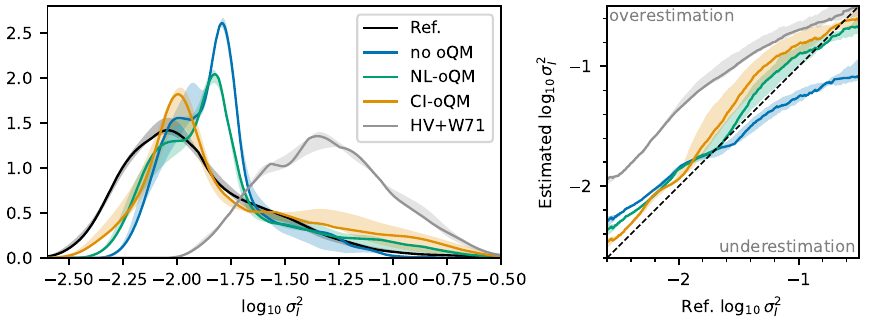}
      \end{minipage}
   \end{subfigure}
   \caption{Canary Islands (out-of-domain test): performance scores, histograms, and quantile-quantile plots for the integrated OT parameters $r_0$ and $\sigma_I^2$ computed from the predicted $C_n^2$ profiles over land. This figure is analogous to fig.~\ref{fig:r0_si2}. Numerical scores are given in tabs.~\ref{tab:app_cn_land_r0} and \ref{tab:app_cn_land_lsi2}.}%
   \label{fig:cn_land_r0_si2}
\end{figure*}

The OTProf model trained on NL is reused without retraining for the Canaries to assess its geographic generalization performance.
We only change the oQM, as this bias-correction explicitly calibrates the DL model outputs to the $C_n^2$ distribution of the training domain.
Table~\ref{tab:cn_experiments} presents the three oQM experiments to be discussed.
ERA5-PL data over the CI serves as input for all three, but the $C_n^2$ output is either not bias-corrected (\emph{no oQM}), bias-corrected using the NL distribution as previously presented (\emph{NL-oQM}), or bias-corrected using a subset of the CI WRF $C_n^2$ data (\emph{CI-oQM}).
The first two experiments use CI data only for validation, but otherwise reflect a situation where no $C_n^2$ reference exists for the new domain.
The third experiment (\emph{CI-oQM}) assumes access to a limited dataset of four 5.5-day simulations over the CI for calibration, with each simulation sampling one season.
The computational effort of ca.~1000~core~hours to run the four simulations is considered generally affordable (cf.~sec.~\ref{sec:wrf} for computational setup).
The remaining eight simulations are held out to quantify the effect of the calibration.
Because calibrating the model output on just 22 days and testing it on just 44 days is noisy, we repeated the procedure three times to assess statistical robustness.
Each iteration uses a different data split, with each calibration subset still sampling all seasons.
The NL analysis did not require this approach because its oQM is fitted on $\sim 180$~days of training data sampled from two different years and tested against the full $\sim 90$~days of test data.

The results are presented in a compressed form compared to the NL domain because most insights carry over, and we only focus on the differences.
For $\log_{10}C_n^2$, randomly drawn profiles and SFs are presented in fig.~\ref{fig:cn_land_lcn2_profiles_sf}, and the scores and PDFs in fig.~\ref{fig:cn_land_lcn2_stats}.
The performance of the integrated parameters is shown in fig.~\ref{fig:cn_land_r0_si2}.

We begin with the vertical structure of the estimated $\log_{10}C_n^2$ profiles, as the complex terrain is the most obvious difference between the CI and the NL training domain.
In the randomly drawn profiles in fig.~\ref{fig:cn_land_lcn2_profiles}, the vertical axis $z$ is the height above ground, so all profiles start at $z=0$ even for elevated locations.
The panels show the terrain height above mean sea level, $z_t$, for each location.
For locations with terrain heights below 1000\,m, the estimated profiles visually match the reference well.
At the location with $z_t=1597$\,m, however, the estimates miss much of the vertical variation of the reference.
This is expected because orographic forcing on atmospheric flow over steep volcanic terrain differs greatly from the flat NL training domain.
Capturing these effects with OTProf would likely require retraining or fine-tuning.
The SFs in fig.~\ref{fig:cn_land_lcn2_sf} show this more clearly.
Here, the solid lines show the median across the three calibration and evaluation splits, and the shaded bands show the minimum and maximum.
At large scales, the slopes of all DL-based SFs are similar to the reference.
However, \emph{no oQM} and \emph{NL-oQM} are offset from the reference by an almost constant factor, indicating too little vertical variability in $\log_{10}C_n^2$.
\emph{CI-oQM} removes this offset for scales larger than about 2000\,m.
At smaller scales, all DL-based SFs deviate from the reference, and \emph{CI-oQM} does not improve them, as calibrating the output distribution cannot add missing vertical structure.
In conclusion, OTProf captures large-scale vertical trends in the new domain but misses fine-scale details, likely shaped by site-specific orographic effects not in the NL training data.

The effect of the oQM becomes more evident in the distributions in fig.~\ref{fig:cn_land_lcn2_stats_dist}.
Without CI calibration, the NL-trained model concentrates too many values around $C_n^2\approx10^{-17}$\,m$^{-2/3}$ and shifts the distribution toward stronger turbulence.
Interestingly, \emph{CI-oQM} recovers the reference distribution almost perfectly, even though the SFs show that the vertical structure is not captured across all scales.
This is possible because the global distribution does not contain information on where in the profile a value occurs.
The temporal and vertical correlations in fig.~\ref{fig:cn_land_lcn2_scores} are lower than in the NL domain and almost identical for all three experiments, as the monotonic oQM hardly changes correlations.
As in the NL domain, part of the low correlation likely stems from differences between the ERA5 and WRF inputs (cf.~fig.~\ref{fig:cn_land_input_agreement}), which are larger for the CI, especially for the surface variables.
WRF resolves the CI at a much finer horizontal resolution than ERA5 (cf.~fig.~\ref{fig:domain_canaries}), so it is expected to reflect the local surface conditions much better.
We consider this an inherent mismatch between the datasets, which likely becomes more problematic in complex terrain than in the flat NL domain.

The lack of fine-scale structure likely also explains the mismatch of the $r_0$ and $\sigma_I^2$ distributions in fig.~\ref{fig:cn_land_r0_si2}.
Without calibration to the CI, $r_0$ is overestimated for most of the range.
This is consistent with the too-low vertical variability in the SFs, as smoothing out thin layers of strong turbulence reduces the integral of $C_n^2$ in linear space.
Surprisingly, \emph{CI-oQM} again brings the $r_0$ distribution much closer to the reference, reducing its errors in skewness and kurtosis.
However, the bias changes sign, so that $r_0$ is now underestimated.
The wide uncertainty bands also reveal the effect of calibrating on only 22~days.
This setup effectively samples one random week per season, making the calibration sensitive to the selected weeks.
For $\sigma_I^2$, the oQM even degrades the performance.
All three experiments overestimate $\sigma_I^2$, but the overestimation grows with \emph{NL-oQM} and even more with \emph{CI-oQM}.
As $r_0$ is most sensitive to near-surface turbulence and $\sigma_I^2$ to upper-air turbulence, this suggests that both oQM variants push upper-air $C_n^2$ too high, while \emph{CI-oQM} improves the near-surface turbulence.
This is consistent with the overestimation of strong scintillation in the NL domain, but more pronounced.
Therefore, we again conclude that the bias correction of the entire column should be refined in future work.

In conclusion, models trained in one domain should not be applied to a very different domain without validation.
However, our results suggest that even a small set of WRF simulations for the new domain, e.g., one week per season, allows benchmarking the performance and improving it through a site-specific oQM.
As the calibration is sensitive to which turbulence conditions are sampled by these few simulations, more simulations, e.g., one week per month, are preferable.
Still, the oQM cannot correct systematic model errors that stem from meteorological conditions not contained in the training data, such as flow over complex orography.
Therefore, future work should explore fine-tuning the pre-trained model to a new domain so it can adapt to such unseen conditions.
We did not pursue fine-tuning in this work because the oQM alone was already sensitive to the selection of calibration and evaluation data.
With only 12 simulations from a single year, we are not confident that conclusions about fine-tuning would be robust.

\section{Conclusion}\label{sec:conclusion}
This study proposed OTProf, a deep learning approach to estimate high-resolution vertical $C_n^2$ profiles (100 levels) from coarse-resolution reanalysis data (30 pressure levels). 
We adapt a variant of the Squeezeformer architecture \citep{kim2022}, designed for atmospheric processes \citep{greysnow2024b}, to also enable vertical super-resolution.
Training data were generated using a year-long mesoscale simulation over the Netherlands at \horres{2}{km} horizontal resolution, with $C_n^2$ computed following the variance-based parameterization of \citet{he2015}.
The simulations were performed using the Weather Research and Forecasting (WRF) model \citep{skamarock2021} and are staggered in time over 4 years to ensure temporal statistical representativeness.
ERA5 reanalysis data \citep{era5} on pressure levels (ERA5-PL) were used as coarse-resolution input data, with input features selected to represent wind shear and buoyancy as the processes that modulate atmospheric turbulence.
Training on datasets from different numerical models (e.g., ERA5-PL and WRF) is problematic because of spatiotemporal misalignment and differences in model resolution and physics parameterizations.
We address this misalignment in two steps.
First, we use a decoupled training approach in which the Squeezeformer is trained on vertically coarsened WRF data (WRF-PL) as input, emulating ERA5-PL.
Second, the trained model is applied to ERA5-PL data, and its outputs are quantile-mapped to the WRF $C_n^2$ distribution.
In Appendix~A, we show that this approach yields more realistic distributions and vertical structure than training directly on ERA5-PL data.
As the reference $C_n^2$ profiles are derived from WRF, all reported scores describe how well OTProf reproduces model-based optical turbulence rather than observations.

Three experiments of increasing complexity were conducted to isolate the contributions of regression, super-resolution, and dataset misalignment to the overall prediction error. 
The simplest experiment (\emph{Ctl.}) uses native high-resolution WRF data for both input and target, the second (\emph{WRF-PL}) introduces super-resolution from 30 to 100 vertical levels using coarsened WRF input, and the third (\emph{ERA5-PL oQM}) applies the trained model to ERA5-PL data to simulate operational conditions.
The analytical Hufnagel-Valley model \citep{hufnagel1974,ulrich1988,valley1980}, commonly used in the optical turbulence community, served as a lower baseline throughout the study.

The results show that the Squeezeformer accurately reconstructs $C_n^2$ profiles in the \emph{Ctl.}\ setting, confirming that the selected meteorological input variables carry sufficient information for this regression task. 
Introducing super-resolution (\emph{WRF-PL}) leads to some smoothing of small-scale features, as quantified by the structure function analysis, but retains the overall vertical structure of the profiles well. 
With ERA5-PL inputs, the output quantile mapping yields $\log_{10}C_n^2$ estimates with almost no bias and a distribution that matches the reference well.
Only the strongest turbulence conditions are underestimated, as in all DL models.
However, the temporal and vertical correlation with the WRF reference is considerably lower for \emph{ERA5-PL oQM} than for the WRF-based experiments.
Comparing ERA5 and WRF inputs suggests that much of this reduction stems from the inherent mismatch between the two datasets.
Because quantile mapping only corrects the distribution, it cannot compensate for differences in the timing and placement of turbulent features.
Future comparisons with observations should assess whether the low correlation is an artifact of comparing mismatched datasets or a limitation of the model.
For now, we recommend using OTProf with ERA5-PL inputs for statistical and climatological assessments, e.g., by temporally extrapolating a limited set of WRF simulations with long ERA5 records.

We assessed practical relevance for FSOC based on the Fried parameter $r_0$ and scintillation index $\sigma_I^2$, both of which improve considerably over the \emph{HV+W71} baseline.
For space-to-ground terminals with aperture diameters between 50 and 100\,cm, the ERA5-based estimates slightly underestimate how often the wavefront is degraded but slightly overestimate how often the degradation is severe.
For $\sigma_I^2$, the estimates match the reference for most conditions and overestimate strong scintillation, leading to conservative link-loss estimates.
However, all DL models underestimate the most extreme turbulence conditions, which can cause communication outages, and this requires further work.

To assess the spatial generalization, we applied the model trained on the Netherlands without retraining to the Canary Islands, which differ strongly in climate and terrain.
Without calibrating to the new domain, the estimated $C_n^2$ distributions and integrated parameters clearly deviate from the reference.
Calibrating the output quantile mapping with only four 5.5-day WRF simulations of the new domain recovers the $C_n^2$ distribution and improves $r_0$.
However, it cannot add the fine-scale structure associated with the complex orography, and it increases the overestimation of $\sigma_I^2$.
Therefore, a trained model should not be applied to a very different domain without validation.
Still, our results suggest that a small set of WRF simulations for the new domain already allows us to benchmark the model and to improve its performance.

OTProf requires an upfront cost for generating the WRF training data, which amounted to ca.\ 23\,500 core hours for one year over the Netherlands.
Afterward, OTProf is computationally cheap.
Training the model took ca.\ 6 GPU hours, and a 30-year $C_n^2$ climatology over the Netherlands could be produced in less than 30\,min of GPU time, whereas simulating it with WRF would require ca.\ 700\,000 core hours.
For a new domain, the out-of-domain results suggest that adapting the pre-trained model with a few local WRF simulations could be considerably cheaper than generating a new year of WRF training data.

Several aspects warrant further investigation. 
The quantile mapping is currently applied to the entire column, but the results for $r_0$ and $\sigma_I^2$ suggest that near-surface and upper-air turbulence require different corrections.
Additionally, fine-tuning the pre-trained model on a small dataset from a new domain could help it adapt to unseen conditions, such as complex orography.
The coarse vertical spacing of ERA5 pressure levels within the boundary layer, where turbulence is typically strongest, is expected to limit the available information in this critical region.
Using ERA5 model levels for training could help, but it would also reduce practicability because pressure-level data are more readily available than model-level data.
Finally, observational data from, e.g., scintillometers ($C_n^2$), scintillation detection and ranging (SCIDAR) instruments ($C_n^2$ profiles), or differential image motion monitors (DIMMs, $r_0$) are needed to validate OTProf experimentally and could further improve the model estimates in the future.
Options range from simple site-specific bias corrections of the estimated $r_0$ using the observed $r_0$ to more complex fine-tuning of the Squeezeformer on observed $C_n^2$ profiles or integrated parameters.
The OTProf architecture supports either option because of grid-agnostic latent-space interpolation, which enables training on various vertical grids, and the existing oQM, which can be repurposed, e.g., to additionally calibrate $r_0$ to observations.

Despite these remaining challenges, OTProf represents a considerable step forward compared to the Hufnagel-Valley model, which remains widely used in the FSOC and optical turbulence communities. 
The HV model is constrained to a fixed exponential profile shape, which cannot capture the diversity of real atmospheric conditions, leading to a persistent positive bias and a poor representation of vertical variability at all but the largest scales. 
OTProf, in contrast, produces profiles that are more realistic in both shape and magnitude while requiring only modest data resources: one year of regional WRF simulations to train and evaluate the model and globally available ERA5 pressure level reanalysis data for inference.
Therefore, we are confident that OTProf offers a promising pathway to more accurate, realistic estimates of vertical $C_n^2$ profiles at manageable computational cost.

\section*{Acknowledgements}
MP is funded by the FREE project (P19-13) of the TTW-Perspectief research program, partially financed by the Dutch Research Council (NWO).
The WRF training dataset was generated using the Dutch national e-infrastructure with the support of the SURF Cooperative using grant no. EINF-15953.
GPU resources for training the deep learning models were provided by University at Albany.

\section*{Data Availability}
The WRF training data and the Python code to train the model will be made available on Zenodo and GitHub, respectively, upon publication.

\bibliographystyle{apsrev4-2}
\bibliography{references}

\appendix

\section{Influence of decoupled training and output quantile mapping}
\label{sec:app_qm}

\begin{figure*}[t]
   \centering
   \begin{subfigure}[t]{\textwidth}
      \centering
      \caption{Randomly drawn examples of $\log_{10} C_n^2$ profiles.}%
      \label{fig:app_lcn2_profiles}
      \includegraphics[width=.8\textwidth]{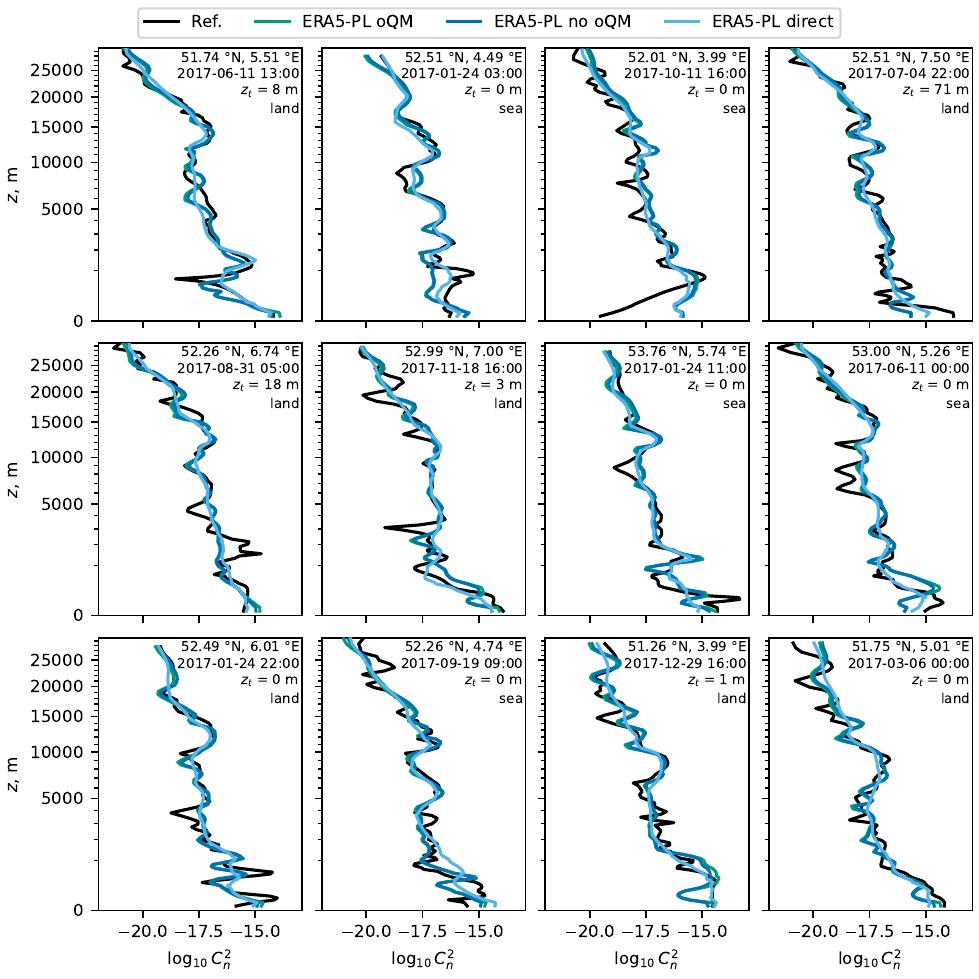}
   \end{subfigure}
   \begin{subfigure}[t]{\textwidth}
      \centering
      \caption{Vertical second-order structure functions of $\log_{10} C_n^2$ profiles.}%
      \label{fig:app_lcn2_sf}
      \includegraphics[width=.9\textwidth]{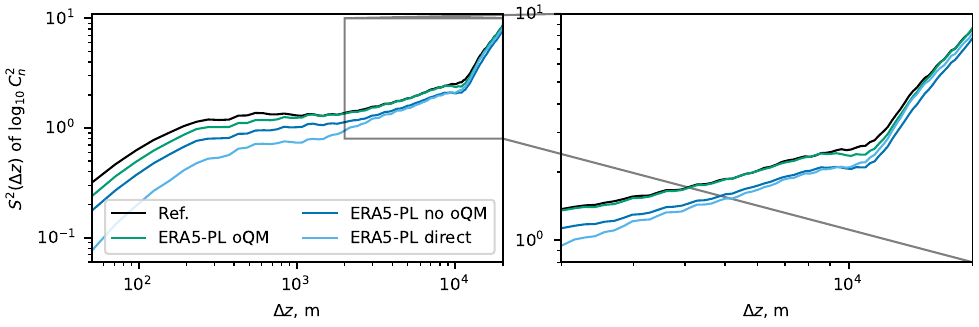}
   \end{subfigure}
   \caption{Examples and vertical structure of predicted $\log_{10} C_n^2$ profiles for three post-processing and training variants. The structure functions are aggregated over all levels, locations, and time. This figure supplements figs.~\ref{fig:lcn2_profiles} and \ref{fig:lcn2_sf}.}%
   \label{fig:app_lcn2_profiles_sf}
\end{figure*}

\begin{figure*}[t]
   \begin{subfigure}[t]{\textwidth}
   \centering
      \caption{$\log_{10} C_n^2$}%
      \label{fig:app_lcn2_stats}
   \begin{minipage}[t]{.35\textwidth}
      \vspace{-\topskip}
      \includegraphics[width=\textwidth]{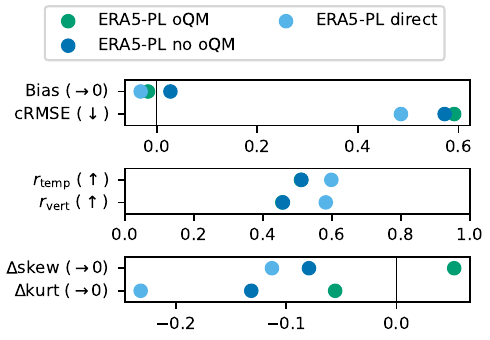}%
   \end{minipage}
   \hfill
   \begin{minipage}[t]{.64\textwidth}
      \vspace{-\topskip}
      \vspace{.4em}
      \includegraphics[width=\textwidth]{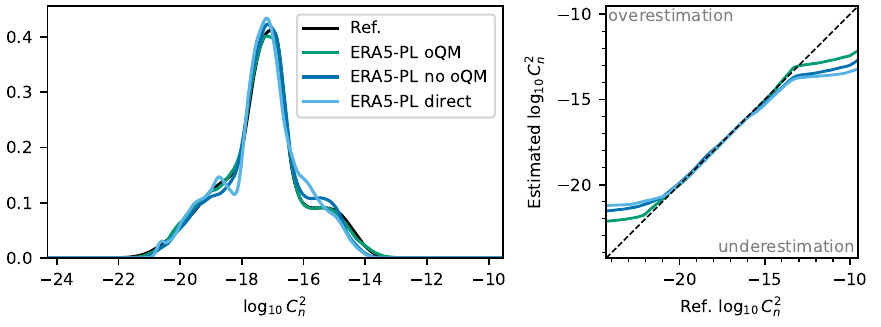}
   \end{minipage}
   \end{subfigure}

   \begin{subfigure}[t]{\textwidth}
   \centering
      \caption{Fried parameter $r_0$}%
      \label{fig:app_r0_stats}
   \begin{minipage}[t]{.35\textwidth}
      \vspace{-\topskip}
      \includegraphics[width=\textwidth]{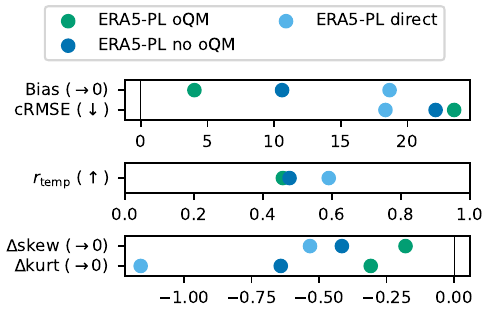}%
   \end{minipage}
   \hfill
   \begin{minipage}[t]{.64\textwidth}
      \vspace{-\topskip}
      \vspace{.4em}
      \includegraphics[width=\textwidth]{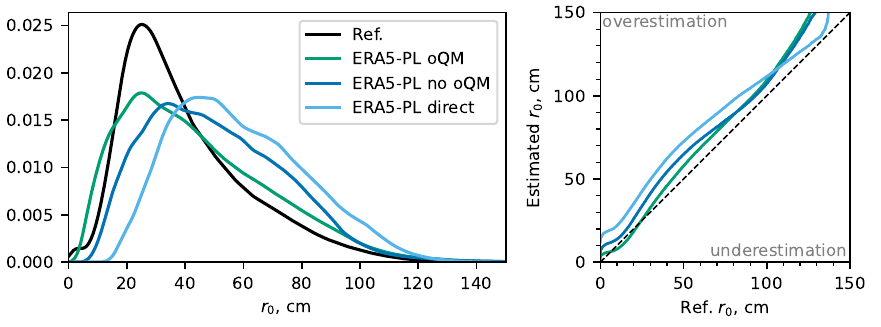}
   \end{minipage}
   \end{subfigure}

   \begin{subfigure}[t]{\textwidth}
      \centering
      \caption{Logarithm of scintillation index $\log_{10}\sigma_I^2$}%
      \label{fig:app_si_stats}
      \begin{minipage}[t]{.35\textwidth}
         \vspace{-\topskip}
         \includegraphics[width=\textwidth]{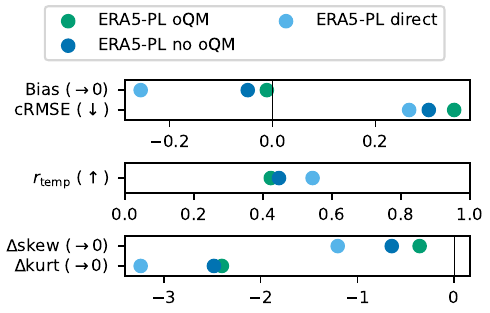}%
      \end{minipage}
      \hfill
      \begin{minipage}[t]{.64\textwidth}
         \vspace{-\topskip}
         \vspace{.4em}
         \includegraphics[width=\textwidth]{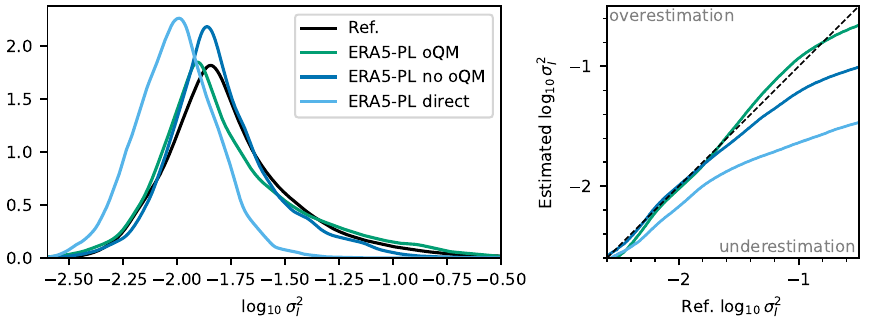}
      \end{minipage}
   \end{subfigure}
   \caption{Performance scores, histograms, and quantile-quantile plots for $\log_{10}C_n^2$ and the integrated OT parameters $r_0$ and $\sigma_I^2$ computed from the predicted $C_n^2$ profiles. Results are shown for three post-processing and training variants. This figure supplements figs.~\ref{fig:lcn2_scores} and \ref{fig:r0_si2}. Numerical scores are given in tabs.~\ref{tab:app_app_lcn2}, \ref{tab:app_app_r0} and \ref{tab:app_app_lsi2}.}%
   \label{fig:app_stats}
\end{figure*}

\begin{table*}[t]
   \caption{Overview of additional experiments to demonstrate the effect of output quantile mapping and decoupled training. Experiment ERA5-PL~oQM (*) is discussed in the main text and repeated here for reference.}%
   \label{tab:app_experiments}
   \begin{tabular}{lcc r@{\:}>{$}c<{$}@{\:}l cc}
      \toprule
      Experiment & \makecell{Regression\\ $f(\mathbf{X}, \mathbf{x}_s; \mathbf{z}_y) \approx \mathbf{Y}$} & \makecell{Superresolution \\ $m \to n$} & \multicolumn{3}{c}{\makecell{Training data\\ $(\mathbf{X}, \mathbf{x}_s) \to \mathbf{Y}$}} & \makecell{Inference data\\ $f(\mathbf{X}, \mathbf{x}_s; \mathbf{z}_y)=\hat{\mathbf{Y}}$} & oQM \\
      \cmidrule(lr){1-1} \cmidrule(lr){2-3} \cmidrule(lr){4-7} \cmidrule(lr){8-8}
      ERA5-PL oQM (*) & \checkmark & \checkmark & WRF-PL & \to & WRF HR & ERA5-PL & \checkmark \\
      ERA5-PL no oQM & \checkmark & \checkmark & \multicolumn{3}{c}{same as above} & ERA5-PL & -- \\
      ERA5-PL direct & \checkmark & \checkmark & ERA5-PL & \to & WRF HR & ERA5-PL & -- \\
      \bottomrule
   \end{tabular}%
\end{table*}

This appendix examines the effect of output quantile mapping (oQM) on the decoupled training approach by comparing three additional experiments summarized in tab.~\ref{tab:app_experiments}. 
The first experiment, \emph{ERA5-PL oQM}, is discussed in the main text as part of the OTProf framework and serves as the reference for the following comparison. 
The second experiment, \emph{ERA5-PL no oQM}, uses the same DL model trained on WRF-PL data but does not apply oQM. 
The third experiment, \emph{ERA5-PL direct}, takes a different approach by directly training a Squeezeformer on ERA5-PL inputs and WRF HR targets, aiming to learn the mapping between the two datasets explicitly rather than relying on decoupled training and statistical alignment. 
The structural differences of the $\log_{10}C_n^2$ profiles predicted by all three methods are presented in fig.~\ref{fig:app_lcn2_profiles_sf}.
The performance scores and statistical characteristics of $\log_{10}C_n^2$, $r_0$, and $\log_{10} \sigma_I^2$ are presented in fig.~\ref{fig:app_stats}.

\paragraph{Output quantile mapping (oQM) versus no correction} 
Comparing \emph{ERA5-PL oQM} with \emph{ERA5-PL no oQM} shows the benefit of statistically adjusting the ERA5-based $C_n^2$ outputs to match WRF's $C_n^2$ distribution.
The structure functions in fig.~\ref{fig:app_lcn2_sf} reveal a clear difference.
Without oQM, the overall vertical variance of ERA5-based $C_n^2$ is too weak, as the constant offset between the \emph{no oQM} curve and the reference indicates.
oQM fixes this shift almost fully for scales larger than a few hundred meters.
A gap remains only at small scales as discussed in the main text.
The $\log_{10}C_n^2$ PDF in fig.~\ref{fig:app_lcn2_stats} responds primarily to oQM, with reduced bias and slightly better-matched tails (lower $\Delta \text{kurt}$).
As quantile mapping is a monotonic map, it preserves the ranks of the values and hardly affects the temporal and vertical correlation of $\log_{10}C_n^2$, which is why the \emph{oQM} and \emph{no oQM} markers almost overlap.
However, the cRMSE increases slightly with oQM.
This is expected because oQM restores the missing variance of the estimates, and for moderate correlation, a larger variance also increases the error amplitude.

However, the distributions of the integrated parameters react more clearly to oQM.
For $r_0$, the bias and the errors in skewness and excess kurtosis are clearly reduced.
For $\sigma_I^2$, oQM reduces the bias and the skewness error, while the excess kurtosis error hardly changes.
As for $\log_{10}C_n^2$, the cRMSE of both parameters increases with oQM.
Since the column integral is nonlinear in the corrected $C_n^2$ values, the temporal correlation of $r_0$ and $\sigma_I^2$ also decreases slightly.
As discussed in the main text, oQM also requires future work to improve the estimates of the OT parameters further, but considering the \emph{no oQM} case, we believe oQM is a step in the right direction.

\paragraph{Decoupled training versus direct training} 
Instead of using quantile mapping to bridge the gap between WRF-PL and ERA5-PL, the \emph{ERA5-PL direct} experiment tests whether training directly on ERA5-PL inputs and WRF HR targets is beneficial. 
This setup lets the model learn a direct mapping between the two datasets, potentially compensating for dataset misalignment during training. 
The increased vertical and temporal correlation scores across all variables presented in fig.~\ref{fig:app_stats} suggest that the direct model can indeed partially compensate for the mismatch.
Direct training also yields a lower cRMSE than the decoupled experiments.
However, all other scores and -- most importantly -- the realistic vertical structure degrade.
The $r_0$ and $\log_{10}\sigma_I^2$ PDFs are biased and do not capture the tails correctly. 
This is likely due to the much wider gap between the reference structure function and \emph{ERA5-PL direct} at small scales in fig.~\ref{fig:app_lcn2_sf}, while at scales above ca.~2\,km, \emph{ERA5-PL direct} is close to \emph{ERA5-PL no oQM}.
The gap reflects much stronger smoothing than the decoupled models, likely missing crucial features relevant to accurate OT parameter estimates.

The comparison of these three experiments shows that output quantile mapping and decoupled training offer a clear advantage in the distribution and vertical structure of the estimates, despite the higher correlation and lower cRMSE in direct training.
For statistical and climatological applications, realistic distributions and vertical structure are more important than the timing of individual events.
As the correlation between ERA5-PL and WRF-PL inputs is already low for some variables (cf.~main text), the low correlations could also be artifacts of the dataset mismatch.
Still, a remaining misalignment is expected, stemming from different physical parameterizations and ERA5's coarser horizontal resolution resolving fewer terrain effects.
Also, oQM seems to overcorrect high-altitude turbulence, leading to too large scintillation indices, as discussed in the main text.
We acknowledge that more work is needed, but based on these findings in this appendix, we chose \emph{ERA5-PL oQM} as the preferred approach for OTProf.
It best balances overall performance with a realistic representation of vertical variability across all scales.

\begin{figure*}[t]
   \includegraphics[width=\textwidth,trim=7mm 6mm 12mm 5mm,clip]{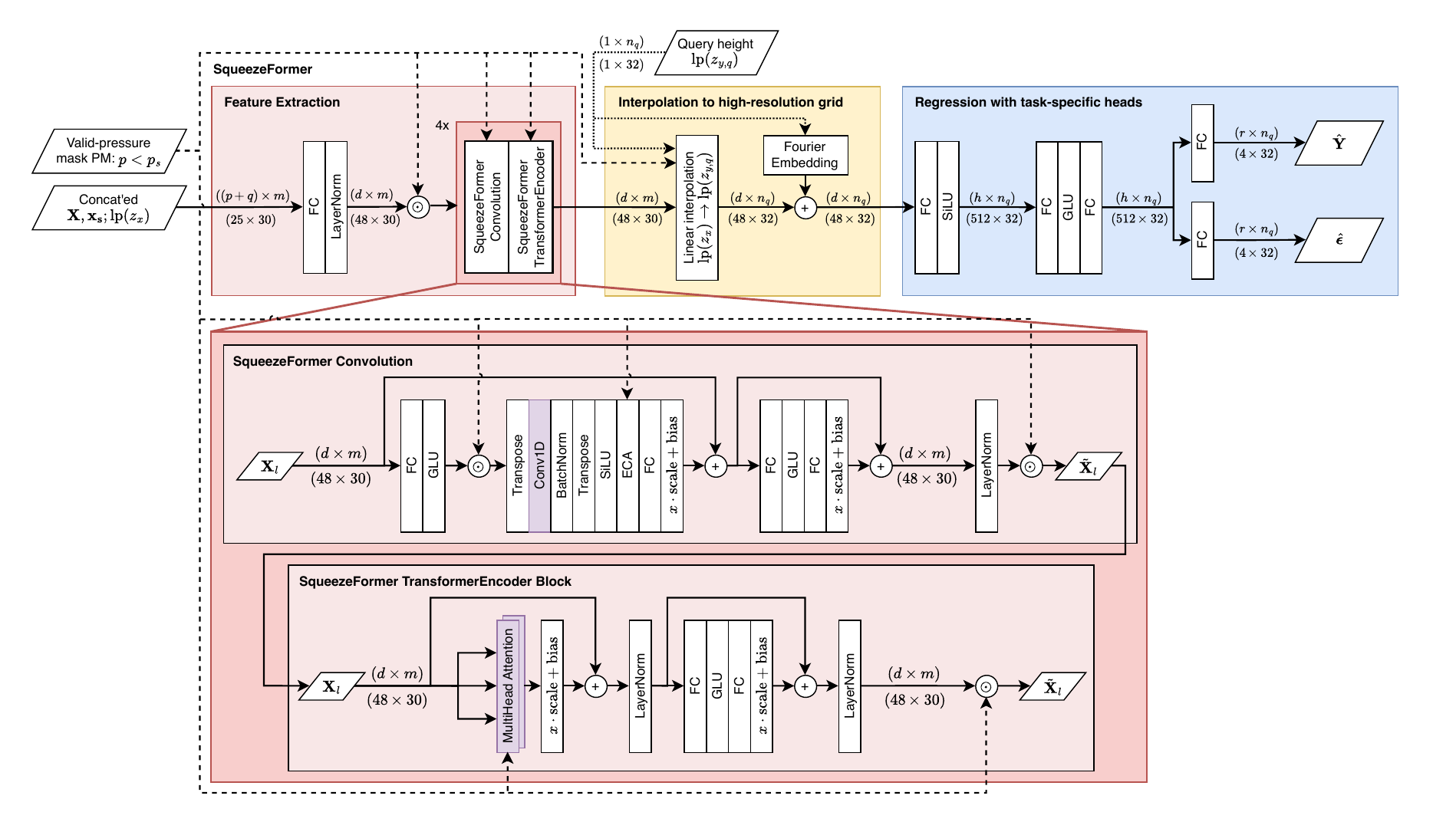}
   \caption{Detailed schematic of the Squeezeformer architecture extending fig.~\ref{fig:sqf_arch} of the main text. Dashed lines indicate where the valid-pressure mask (PM) enters.}\label{fig:sqf_arch_detail}
\end{figure*}

\section{Additional details of Squeezeformer architecture}
\label{sec:app_sqf_arch}

Figure~\ref{fig:sqf_arch_detail} provides a more detailed schematic of the Squeezeformer architecture compared to fig.~\ref{fig:sqf_arch} presented in the main text.
Primarily, the figure illustrates the composition of the Squeezeformer convolution and transformer blocks, and the locations where the valid-pressure mask (PM) enters.
The convolution block takes the embedded latent input $\mathbf{X}_l$ and expands the number of channels by a factor of 4 with an FC layer, which is halved again by the GLU activation, before a depthwise convolution is applied.
To avoid signals from below-ground pressure levels, element-wise multiplication of the expanded channels with the PM cancels invalid information before the convolutions.
Efficient channel attention (ECA, \citet{wang2020}) is applied to capture cross-channel interactions followed by a projection back to the original number of channels.
The ECA layer also receives the PM to exclude below-ground levels from the channel-wise average.
Learnable scaling and bias parameters are applied per channel before adding $\mathbf{X}_l$ via a residual connection.
Another GLU-activated embedding block is applied, followed by learnable scaling and bias parameters, then a second residual connection and layer normalization.
The final output of the Squeezeformer convolution block is again masked with PM for consistency.
GLU activations throughout the architecture allow the network to suppress irrelevant information and focus on important features.

The convolution block output is fed into the transformer block.
Two attention heads are used to capture interactions across the vertical dimension, followed by learnable scaling and bias parameters.
The PM enters the attention heads as an attention mask to avoid attending to below-ground values.
The output is normalized after adding the residual connection and then embedded through a GLU-activated fully connected network, as in the convolution block.
After adding a second residual connection, we apply another LayerNorm and PM to yield the transformer block's final output.

After multiple Squeezeformer blocks, linear interpolation in the latent space is performed on the log-scaled height coordinate shared between ERA5 and WRF (cf.~eq.~\eqref{eq:lz}).
To provide vertical positional encoding to the downstream regression decoder, we add a Fourier embedding of the scaled query heights to the interpolated latent features.
The Fourier embedding represents each scaled query height $\textrm{lp}(z)$ by sine and cosine functions, $\sin(\pi f_k\, \textrm{lp}(z))$ and $\cos(\pi f_k\, \textrm{lp}(z))$, at 16 frequencies $f_k$ that are logarithmically spaced between 1 and 64.
Low frequencies vary slowly across the column and encode the coarse vertical position, while high frequencies vary quickly and allow the model to distinguish closely spaced levels.
The resulting 32 features are projected by a linear layer to the latent dimension $d=48$ before they are added to the interpolated latent features.

\section{Tabulated performance scores}
For clarity, the score figures in the main text do not show the numerical values of the different metrics, but they are given in the tables below.

\begin{table*}
   \caption{Performance scores of the in-domain experiments shown in figs.~\ref{fig:lcn2_scores} and \ref{fig:r0_si2}.}\label{tab:app_main}
   \centering
   \begin{subtable}{.48\textwidth}
      \centering
      \caption{$\log_{10}C_n^2$}\label{tab:app_main_lcn2}
      \begin{tabular}{lcccc}
\toprule
 & Ctl. & WRF-PL & \makecell{ERA5-PL\\oQM} & HV+W71 \\
\midrule
Bias ($\to 0$) & -0.000 & -0.010 & -0.017 & 0.584 \\
cRMSE ($\downarrow$) & 0.142 & 0.298 & 0.592 & 0.656 \\
$r_\mathrm{temp}$ ($\uparrow$) & 0.962 & 0.854 & 0.510 & 0.193 \\
$r_\mathrm{vert}$ ($\uparrow$) & 0.962 & 0.858 & 0.456 & 0.119 \\
$\Delta$skew ($\to 0$) & -0.022 & 0.021 & 0.052 & -0.342 \\
$\Delta$kurt ($\to 0$) & -0.013 & -0.160 & -0.056 & 0.587 \\
\bottomrule
\end{tabular}

   \end{subtable}

   \vspace{1em}
   \begin{subtable}{.48\textwidth}
      \centering
      \caption{$r_0$ in cm}\label{tab:app_main_r0}
      \begin{tabular}{lcccc}
\toprule
 & Ctl. & WRF-PL & \makecell{ERA5-PL\\oQM} & HV+W71 \\
\midrule
Bias ($\to 0$) & 1.727 & 4.629 & 4.027 & -13.561 \\
cRMSE ($\downarrow$) & 8.751 & 12.512 & 23.494 & 22.105 \\
$r_\mathrm{temp}$ ($\uparrow$) & 0.897 & 0.800 & 0.458 & 0.163 \\
$\Delta$skew ($\to 0$) & -0.030 & -0.088 & -0.180 & -0.837 \\
$\Delta$kurt ($\to 0$) & -0.172 & -0.191 & -0.309 & -1.947 \\
\bottomrule
\end{tabular}

   \end{subtable}
   \begin{subtable}{.48\textwidth}
      \centering
      \caption{$\log_{10}\sigma_I^2$}\label{tab:app_main_lsi2}
      \begin{tabular}{lcccc}
\toprule
 & Ctl. & WRF-PL & \makecell{ERA5-PL\\oQM} & HV+W71 \\
\midrule
Bias ($\to 0$) & -0.048 & -0.091 & -0.011 & 0.286 \\
cRMSE ($\downarrow$) & 0.180 & 0.212 & 0.353 & 0.344 \\
$r_\mathrm{temp}$ ($\uparrow$) & 0.820 & 0.742 & 0.423 & 0.224 \\
$\Delta$skew ($\to 0$) & -0.675 & -0.734 & -0.357 & -1.141 \\
$\Delta$kurt ($\to 0$) & -2.558 & -2.451 & -2.400 & -3.794 \\
\bottomrule
\end{tabular}

   \end{subtable}
\end{table*}

\begin{table*}
   \caption{Performance scores of the out-of-domain experiments (land only) shown in figs.~\ref{fig:cn_land_lcn2_stats} and \ref{fig:cn_land_r0_si2}.}\label{tab:app_cn_land}
   \centering
   \begin{subtable}{.75\textwidth}
      \centering
      \caption{$\log_{10}C_n^2$}\label{tab:app_cn_land_lcn2}
      \begin{tabular}{lcccc}
\toprule
 & \makecell{no\\oQM} & NL-oQM & CI-oQM & HV+W71 \\
\midrule
Bias ($\to 0$) & 0.142 $\pm$ 0.006 & 0.096 $\pm$ 0.008 & 0.000 $\pm$ 0.025 & 0.697 $\pm$ 0.011 \\
cRMSE ($\downarrow$) & 0.557 $\pm$ 0.001 & 0.572 $\pm$ 0.001 & 0.607 $\pm$ 0.004 & 0.610 $\pm$ 0.004 \\
$r_\mathrm{temp}$ ($\uparrow$) & 0.370 $\pm$ 0.003 & 0.369 $\pm$ 0.003 & 0.371 $\pm$ 0.003 & 0.228 $\pm$ 0.013 \\
$r_\mathrm{vert}$ ($\uparrow$) & 0.288 $\pm$ 0.004 & 0.285 $\pm$ 0.005 & 0.284 $\pm$ 0.007 & 0.136 $\pm$ 0.014 \\
$\Delta$skew ($\to 0$) & -0.212 $\pm$ 0.054 & -0.057 $\pm$ 0.063 & -0.012 $\pm$ 0.115 & -0.126 $\pm$ 0.027 \\
$\Delta$kurt ($\to 0$) & 0.596 $\pm$ 0.093 & 0.802 $\pm$ 0.098 & 0.019 $\pm$ 0.150 & -0.153 $\pm$ 0.045 \\
\bottomrule
\end{tabular}

   \end{subtable}

   \vspace{1em}
   \begin{subtable}{.75\textwidth}
      \centering
      \caption{$r_0$ in cm}\label{tab:app_cn_land_r0}
      \begin{tabular}{lcccc}
\toprule
 & \makecell{no\\oQM} & NL-oQM & CI-oQM & HV+W71 \\
\midrule
Bias ($\to 0$) & 12.051 $\pm$ 2.145 & 6.684 $\pm$ 2.656 & -6.989 $\pm$ 4.531 & -33.946 $\pm$ 2.264 \\
cRMSE ($\downarrow$) & 33.031 $\pm$ 0.708 & 35.806 $\pm$ 0.861 & 32.468 $\pm$ 1.715 & 31.671 $\pm$ 1.103 \\
$r_\mathrm{temp}$ ($\uparrow$) & 0.312 $\pm$ 0.035 & 0.290 $\pm$ 0.036 & 0.307 $\pm$ 0.040 & 0.064 $\pm$ 0.026 \\
$\Delta$skew ($\to 0$) & -0.767 $\pm$ 0.051 & -0.565 $\pm$ 0.072 & -0.084 $\pm$ 0.159 & 0.846 $\pm$ 0.220 \\
$\Delta$kurt ($\to 0$) & -1.090 $\pm$ 0.272 & -1.080 $\pm$ 0.160 & 0.143 $\pm$ 0.403 & 2.940 $\pm$ 0.866 \\
\bottomrule
\end{tabular}

   \end{subtable}

   \vspace{1em}
   \begin{subtable}{.75\textwidth}
      \centering
      \caption{$\log_{10}\sigma_I^2$}\label{tab:app_cn_land_lsi2}
      \begin{tabular}{lcccc}
\toprule
 & \makecell{no\\oQM} & NL-oQM & CI-oQM & HV+W71 \\
\midrule
Bias ($\to 0$) & 0.074 $\pm$ 0.015 & 0.105 $\pm$ 0.031 & 0.142 $\pm$ 0.069 & 0.600 $\pm$ 0.036 \\
cRMSE ($\downarrow$) & 0.328 $\pm$ 0.009 & 0.378 $\pm$ 0.012 & 0.407 $\pm$ 0.011 & 0.403 $\pm$ 0.001 \\
$r_\mathrm{temp}$ ($\uparrow$) & 0.287 $\pm$ 0.046 & 0.264 $\pm$ 0.045 & 0.296 $\pm$ 0.050 & 0.148 $\pm$ 0.028 \\
$\Delta$skew ($\to 0$) & -0.446 $\pm$ 0.056 & -0.115 $\pm$ 0.151 & -0.490 $\pm$ 0.283 & -1.007 $\pm$ 0.086 \\
$\Delta$kurt ($\to 0$) & -1.692 $\pm$ 0.716 & -1.389 $\pm$ 0.990 & -2.984 $\pm$ 1.003 & -3.149 $\pm$ 0.316 \\
\bottomrule
\end{tabular}

   \end{subtable}
\end{table*}

\begin{table*}
   \caption{Performance scores of the Appendix~A experiments shown in fig.~\ref{fig:app_stats}.}\label{tab:app_app}
   \centering
   \begin{subtable}{.48\textwidth}
      \centering
      \caption{$\log_{10}C_n^2$}\label{tab:app_app_lcn2}
      \begin{tabular}{lccc}
\toprule
 & \makecell{ERA5-PL\\oQM} & \makecell{ERA5-PL\\no oQM} & \makecell{ERA5-PL\\direct} \\
\midrule
Bias ($\to 0$) & -0.017 & 0.027 & -0.032 \\
cRMSE ($\downarrow$) & 0.592 & 0.573 & 0.486 \\
$r_\mathrm{temp}$ ($\uparrow$) & 0.510 & 0.512 & 0.598 \\
$r_\mathrm{vert}$ ($\uparrow$) & 0.456 & 0.458 & 0.582 \\
$\Delta$skew ($\to 0$) & 0.052 & -0.079 & -0.113 \\
$\Delta$kurt ($\to 0$) & -0.056 & -0.132 & -0.232 \\
\bottomrule
\end{tabular}

   \end{subtable}

   \vspace{1em}
   \begin{subtable}{.48\textwidth}
      \centering
      \caption{$r_0$ in cm}\label{tab:app_app_r0}
      \begin{tabular}{lccc}
\toprule
 & \makecell{ERA5-PL\\oQM} & \makecell{ERA5-PL\\no oQM} & \makecell{ERA5-PL\\direct} \\
\midrule
Bias ($\to 0$) & 4.027 & 10.597 & 18.651 \\
cRMSE ($\downarrow$) & 23.494 & 22.101 & 18.344 \\
$r_\mathrm{temp}$ ($\uparrow$) & 0.458 & 0.477 & 0.591 \\
$\Delta$skew ($\to 0$) & -0.180 & -0.415 & -0.533 \\
$\Delta$kurt ($\to 0$) & -0.309 & -0.641 & -1.160 \\
\bottomrule
\end{tabular}

   \end{subtable}
   \begin{subtable}{.48\textwidth}
      \centering
      \caption{$\log_{10}\sigma_I^2$}\label{tab:app_app_lsi2}
      \begin{tabular}{lccc}
\toprule
 & \makecell{ERA5-PL\\oQM} & \makecell{ERA5-PL\\no oQM} & \makecell{ERA5-PL\\direct} \\
\midrule
Bias ($\to 0$) & -0.011 & -0.048 & -0.256 \\
cRMSE ($\downarrow$) & 0.353 & 0.304 & 0.266 \\
$r_\mathrm{temp}$ ($\uparrow$) & 0.423 & 0.447 & 0.544 \\
$\Delta$skew ($\to 0$) & -0.357 & -0.645 & -1.203 \\
$\Delta$kurt ($\to 0$) & -2.400 & -2.481 & -3.238 \\
\bottomrule
\end{tabular}

   \end{subtable}
\end{table*}

\end{document}